\documentclass[prb,twocolumn,amsmath,amssymb,amsbsy,superscriptaddress,floatfix,noshowpacs,10pt,longbibliography]{revtex4-1}\usepackage{graphicx,bm,times}
\usepackage{amsmath}
\usepackage{amsfonts}
\usepackage{amssymb}
\usepackage{color}

\begin{document}

\title{Modelling the angle-dependent magnetoresistance oscillations of \\
Fermi surfaces with hexagonal symmetry}

\author{Joseph C. A. Prentice}
\affiliation{Clarendon Laboratory, Department of Physics,
University of Oxford, Parks Road, Oxford, OX1 3PU, U.K.}
\affiliation{TCM Group, Cavendish Laboratory, University of Cambridge,
 J. J. Thomson Avenue, Cambridge, CB3 0HE, U.K.}
\author{Amalia I. Coldea}
\affiliation{Clarendon Laboratory, Department of Physics,
University of Oxford, Parks Road, Oxford, OX1 3PU, U.K.}

\begin{abstract}
By solving the Boltzmann transport equation we investigate theoretically the general form of oscillations in the resistivity caused by varying the direction of an applied magnetic field for the case of quasi-two dimensional systems on hexagonal lattices.
The presence of the angular magnetoresistance oscillations can be used to map out the topology of the Fermi surface and we study how this effect varies as a function of the degree of inter-plane warping as well as a function of the degree of isotropic scattering. We find that the angular dependent effect due to in-plane rotation follows the symmetry imposed by the lattice whereas for inter-plane rotation the degree of warping dictates the dominant features observed in simulations.
Our calculations make predictions for specific angle-dependent magnetotransport
signatures in magnetic fields expected for quasi-two dimensional hexagonal compounds similar to PdCoO$_2$ and PtCoO$_2$.
\end{abstract}
\today
\maketitle

\section{Introduction}
The electronic and thermal properties of exotic metals and superconductors originate from the intricate details of their Fermi surfaces. Understanding the Fermi surface of a material requires direct experimental measurements through various techniques, such as angle-resolved photoemission spectroscopy (ARPES) \cite{Damascelli} and quantum oscillations experiments by measuring the de Haas-van Alphen (dHvA) and/or Shubnikov-de Haas (SdH) effects \cite{Ashcroft}. Another powerful method of understanding the shape of simple Fermi surfaces is angle-dependent magnetoresistance oscillations (AMRO), which allow access to the Fermi surface at much higher temperatures and scattering rates than the purely quantum oscillation effects. This method has been employed successfully to map both quasi-one dimensional (Q1D) and quasi-two dimensional (Q2D) Fermi surfaces, as in BEDT-TTF organic salts \cite{Goddard,Singleton}, intercalated graphite compounds \cite{Enomoto}, ruthenates \cite{Bergemann}, and superconducting pnictides \cite{Kimata} and cuprates \cite{Abdel-Jawad,Hussey,Analytis,Kang,Lewin,Nowojewski_2008,Nowojewski_2010}.

Many electronic systems with hexagonal symmetry possess interesting transport properties, such as unusual anisotropy in electronic scattering, charge-density waves phenomena and backscattering protected transport on the surface of topological insulators. PdCoO$_2$ is one such interesting material, belonging to a more general family of delafossite compounds, which has been found to have extremely large magnetoresistance \cite{Takatsu} and unusual transport properties \cite{Ong}, with strong transport anisotropy between the in-plane and out-of-plane directions, which differ by a factor of up to 200 \cite{Eyert}. Recent experimental angle-dependent studies show very strong features in the magnetoresistance of PdCoO$_2$, both when the magnetic field is rotated
in the conductive plane \cite{Takatsu} as well as out-of-plane \cite{Kikugawa}. Furthermore, quantum oscillations show that
the data can be modelled using a single corrugated hexagonal Fermi surface \cite{Hicks}.

In this work, we use the Boltzmann equation\cite{Ziman} to calculate the angle dependent magnetotransport properties
for electronic systems with hexagonal symmetry, which have not been explored before. Starting from a tight-binding description of the Fermi surface on a hexagonal lattice, we investigate the angle dependent magnetoresistance when the magnetic field is rotated, either in the conducting plane or out-of-plane, focusing in particular on the role of scattering and
the degree of warping of the Fermi surface. We find that the form of the magnetoresistance depends strongly on the degree of warping, with additional features emerging at high warping levels. Our work provides a large range of parameters
 that can be used to compare with future experimental studies for materials with a single quasi-two dimensional hexagonal Fermi surface, such as PdCoO$_2$ and PtCoO$_2$.

\section{Fermi surface harmonic expansion}

\begin{figure*}[htbp]
\centering
\includegraphics[width=\textwidth,clip=true]{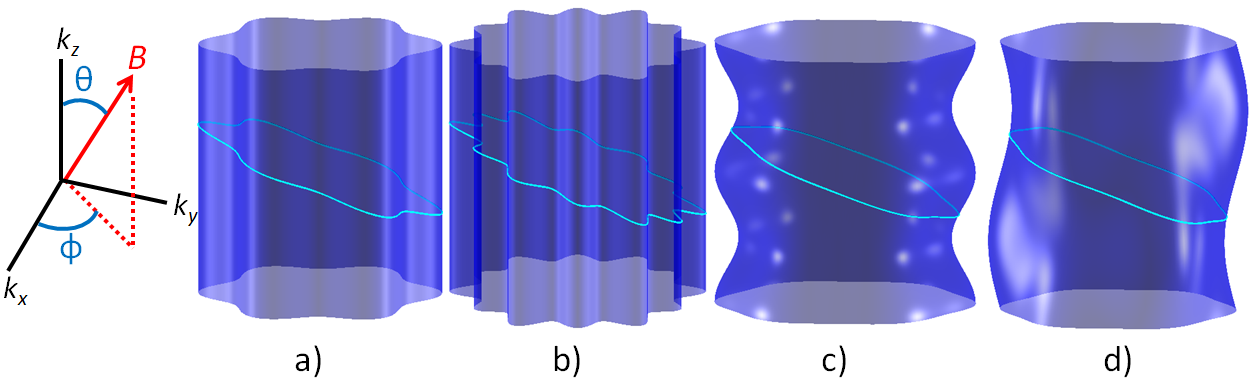}
\caption{(colour online) Simulated hexagonal Fermi surfaces. The effect of different anisotropic terms
defined in Eq. \eqref{eq:kfexpansion} and set as $k_{00}$=0.9559 throughout
and a) $k_{0,6}$=0.1, b)  $k_{0,12}$=0.1 and $k_{0,6}$=0.04,
c) $k_{2,0}$=0.1 and $k_{0,6}$=0.04 and d) $k_{1,3}$=0.1 and $k_{0,6}$=0.04, all
parameters in units of $\text{\r{A}}^{-1}$.
Solid line indicates a typical cyclotron orbit for a particular Fermi
surface when the applied magnetic field makes the polar angle, $\theta$, and azimuthal angle, $\phi$, with respect to
the direction of the magnetic field $\mathbf{B}$, as shown in the left panel.}
\label{fig:CoefficientsFS}
\end{figure*}

In order to investigate analytically a single Fermi surface on a hexagonal lattice
 we used a tight-binding approach to expand a quasi-two dimensional Fermi surface
 using cylindrical harmonics, as considered in previous studies \cite{Hicks,Analytis,Bergemann}:

\begin{equation} k_F (\psi,k_z) = \sum_{m,n \geq0} k_{n,m} \left \{
\begin{array}{l}
\cos \\
\sin
\end{array}
\right \} m\psi \times \left \{
\begin{array}{l}
\cos \\
\sin
\end{array}
\right \} n\kappa \end{equation}
where $\kappa = dk_z = \frac{ck_z}{3}$, $d$ is the spacing between conducting layers and $\psi$ is the azimuthal angle in the cylindrical co-ordinate system used. $\kappa \in [ -\pi, \pi]$. $k_{00}$ gives the average radius of the FS, whilst the other coefficients give various corrugations and warpings of the surface, as shown in Fig. \ref{fig:CoefficientsFS}. This expansion must obey the symmetries of the hexagonal lattice as identified by the following symmetry operations:
\begin{enumerate}
\item Rotation and translation: $\psi \to \psi + \frac{\pi}{3}, \kappa \to \kappa + \pi$
\item Rotation: $\psi \to \psi + \frac{2\pi}{3}$
\item Inversion: $\psi \to \psi + \pi, \kappa \to -\kappa$
\item Reflection 1: $\psi \to -\psi$
\item Reflection 2: $\psi \to \frac{2\pi}{3} - \psi$
\end{enumerate}
suggesting that $m$ and $n$ must both be odd or both be even based on 1), $m \bmod 3 = 0$ based on 2), $\cos n\kappa$ terms must be accompanied by even $m$ and $\sin n\kappa$ terms must be accompanied by odd $m$ based on 3), and only $\cos m\psi$ terms are allowed based on 4) and 5).  These operations are discussed in more detail in Appendix J. Taking all these constraints and assuming that near-neighbour hopping dominates by ignoring higher order terms of the expansion \cite{Grigoriev}, we are left with the expression:
\begin{equation} \label{eq:kfexpansion}
\begin{aligned} k_F &= k_{00} + k_{0,6} \cos 6\psi + k_{0,12} \cos 12\psi \\ &+ k_{2,0} \cos 2\kappa + k_{1,3} \sin \kappa \cos 3\psi . \end{aligned}
\end{equation}
Fig.\ref{fig:CoefficientsFS} shows this for different values of the coefficients. This expression is similar to that reported in Ref. \onlinecite{Hicks}, except the additional term, $k_{0,1}$, which is ruled out by the first symmetry operation. Physically, the $k_{2,0}$ and $k_{1,3}$ terms are related to interlayer anisotropy of the Fermi surface, whilst the other terms are related to in-plane anisotropy. The effect of each of these anisotropic terms on the shape of the Fermi surface is shown in Fig. \ref{fig:CoefficientsFS}, where $k_{0,6}$ introduces a 6-fold symmetry whilst $k_{0,12}$ introduces a 12-fold symmetry, and $k_{2,0}$ and $k_{1,3}$ terms introduce warping in the $k_z$-direction. In this work $k_{1,3}$ will be used as a variable to investigate the effect of FS warping on AMRO and the values of the parameters are chosen to be $k_{00}=0.9559, k_{0,6}=0.04, k_{0,12}=0.007$ and $k_{2,0}=-0.0025$~(in units of $\text{\r{A}}^{-1}$) taken to match closely the experimental values for PdCoO$_2$,
as determined from quantum oscillations \cite{Hicks}. We choose $k_{1,3}$ as our variable parameter because we expect it will give a rich structure to the resulting AMRO - as we are calculating the $c$-axis resistivity, we would expect terms containing $k_z$ dependence to have the largest effect on the AMRO. Using $k_{1,3}$ as a variable allows us to explore a range of plausible AMRO resulting from warped PdCoO$_2$-like Fermi surfaces, whereas using other parameters from the expansion would affect the AMRO in a more subtle way.

These parameters can be directly related to the transfer integrals in a usual tight-binding model as detailed in Refs. \onlinecite{Takatsu,Hicks} and discussed in Appendix I. However, when directly inferring the tight-binding overlap integrals from quantum oscillation experiments there are certain limitations - the bare electronic bandwidth is inaccessible by these experiments, and only the states near the Fermi surface are probed, meaning one can determine the Fermi surface geometry and the renormalized bandwidth. The energy bands and the transfer integrals lose their meaning away from the Fermi surface in a strongly interacting system. Thus, the use of a cylindrical harmonic expansion allows us to parametrize the Fermi surface geometry directly from the electronic structure information obtained from experimental data, in particular when comparing quantum oscillations and AMRO data. 

In order to numerically calculate the AMRO, for fields both in and out of the ($ab$)-plane, we calculate the conductivity, and thus $\rho_{zz}$, for a certain FS of the form in Eq. \eqref{eq:kfexpansion}, using MATLAB\cite{MATLAB}.
In order to calculate the AMRO in this work, we calculated $\rho_{zz}$ simply by taking $\rho_{zz}=\frac{1}{\sigma_{zz}}$, similar to previous work \cite{Analytis}.  This expression follows from the fact that off-diagonal terms in the conductivity matrix are generally much smaller than the diagonal ones, as they directly depend on the warping of the Fermi surface, and those off-diagonal elements containing a $z$-component are even smaller, as $v_z \ll v_x ,v_y$, as shown in Appendix E. Results using this expression agree well with further calculations done using the more exact method of calculating and inverting the whole conductivity matrix to obtain $\rho_{zz}$, as shown in Fig.~\ref{fig:ComparingSimplifiedCalcs} in Appendix F, and were also less prone to numerical errors.

 Throughout this work, anisotropy in the effective mass of the electrons and the scattering time is neglected, with
 the values of the isotropic scattering time taken as $\tau=10$ ps and the quasiparticle effective mass as $m^*=1.5m_e$
(after Refs. \onlinecite{Hicks,Takatsu}).  Previous calculations of AMRO without
the inclusion of anisotropy in $\tau$ and $\omega_c = \frac{eB}{m^*}$ have also been shown to reproduce the key features of the AMRO seen in experiments, such as in previous work on quasi-two-dimensional cuprates \cite{Hussey}.
The AMRO features are captured qualitatively by the isotropic calculations, with differences starting to emerge at large $\theta$ angles \cite{Kennett}. This implies that we can draw valid conclusions from calculations that do not include anisotropy in the scattering time or effective mass. This conclusion is also justified by experiments on PdCoO$_2$ up to $100$~K, which are described by a single isotropic scattering rate \cite{Takatsu}, implying that impurity scattering is the dominant process in the regime in which AMRO are observed. In general, anisotropy in the effective mass, and thus the cyclotron frequency $\omega_c$, arises via the expression $\frac{\mathbf{k}_F \cdot \mathbf{v}_F}{k_F^2}$, where $\mathbf{k}_F$ points along the cylindrical radial direction \cite{Kennett}. Anisotropy arises if the two vectors are not parallel. From the results shown in Appendix E, the angle between $\mathbf{k}_F$ and $\mathbf{v}_F$ will be $\arctan \sqrt{(\frac{1}{k_F} \frac{\partial k_F}{\partial \psi})^2 + (\frac{\partial k_F}{\partial k_z})^2}$, and thus the dot product will be proportional to the cosine of this quantity. As the argument of the inverse tangent is small, we can expand the cosine as $1-(\frac{1}{k_F} \frac{\partial k_F}{\partial \psi})^2 - (\frac{\partial k_F}{\partial k_z})^2$, meaning the anisotropy in $m^*$ and $\omega_c$ is small, and will be a second order effect at best, justifying its neglect. Anisotropies in the scattering time or effective mass may introduce important quantitative corrections to our results \cite{Analytis}, but this is beyond the scope of the current work.

The oscillations in AMRO are semi-classical in nature, and are due to the formation of cyclotron orbits on the FS and the changing area enclosed by them \cite{Yamaji}. AMRO can be calculated from a linearised Boltzmann transport equation \cite{Ziman}, which is derived in full in Appendix A. In order to calculate the magnetoresistance and thus observe AMRO, we must derive an expression for the conductivity tensor for the FS given by Eq. \eqref{eq:kfexpansion}. We used the Boltzmann equation:
\begin{equation} \label{eq:Boltzmannequation}
e \mathbf{E} \cdot \mathbf{v} \left(- \frac{\partial f^0_\mathbf{k}}{\partial \varepsilon}\right) = \frac{g_\mathbf{k}}{\tau} + \frac{e}{\hbar}(\mathbf{v}\times\mathbf{B}) \frac{\partial g_\mathbf{k}}{\partial \mathbf{k}}
\end{equation}
where $f^0_\mathbf{k}$ is the particle distribution in the absence of fields or temperature gradients, $g_\mathbf{k}$ is the difference between the steady state distribution and $f^0_\mathbf{k}$ (assumed to be small), $\mathbf{v}$ is the velocity of the particles and $\tau$ is the scattering time. Solving this equation and using it to calculate the conductivity produces the Shockley-Chambers tube integral\cite{Shockley,Chambers}:
\begin{align} \label{eq:SCTI} \sigma_{ij} = \frac{e^2}{4\pi^3\hbar^2} &\int dk_B \int_0^{2\pi} d\alpha \int_0^{\infty} d\alpha'' \nonumber\\ &v_i (\alpha) v_j (\alpha - \alpha'') \frac{m^*}{\omega_c} e^{-\frac{\alpha''}{\omega_c \tau}} . \end{align}
This is derived in detail, in the case of closed orbits only, in the Appendices.

A generalised version of the Shockley-Chambers tube integral that includes the effect of open orbits is \cite{Analytis,Blundell}
\begin{equation} \label{eq:GeneralisedSCTI} \sigma_{ij} = \frac{e^2}{4\pi^3} \int d^3 \mathbf{k} \left(-\frac{\partial f_\mathbf{k}^0}{\partial \varepsilon}\right) v_i (\mathbf{k},0) \int_{-\infty}^0 v_j (\mathbf{k},t) e^{\frac{t}{\tau}} dt .\end{equation}
The time integral integrates over the history of the open orbit.
Here we consider two cases: when the magnetic field is within the $(ab)$-plane ($|\theta|=90^\circ$)
 and the contribution from the open orbits is dominant, or
 when the field is away from the $(ab)$-plane ($|\theta|\leq70^\circ$)
and the closed orbits dominate.

The Shockley-Chambers tube integral gives us an expression for the conductivity tensor, $\sigma_{ij}$, in terms of $k_{\rm F}$ and related quantities such as $v_F$ and $v_z$. However, such an expression does not explicitly show that the magnetoresistance is oscillatory as the direction of the magnetic field varies. However,  previous work \cite{McKenzie,Yamaji,Yagi} has shown that, in the limit of $td \tan \theta \ll \hbar v_F$, where $t$ is the interlayer transfer integral and $d=\frac{c}{3}$, this behaviour is general and allows us to find an expression for the positions of the peaks in the magnetoresistance \cite{Singleton}. For a simple cylindrical Fermi surface with no in-plane warping, the inter-plane component of the conductivity, $\sigma_{zz}$ takes the form \cite{McKenzie,Yagi} \begin{equation} \label{eq:McKenzieForm} \frac{\sigma_{zz}(\theta)}{\sigma_{zz}(0)} = [J_0 (\mu) ]^2 + 2 \sum_{\nu=1}^{\infty} \frac{[J_\nu (\mu)]^2}{1+(\nu \omega_c \tau \cos \theta)^2}. \end{equation}
$\mu = dk_{\parallel} \tan \theta$, where $k_{\parallel}$ is the maximum possible projection of an in-plane Fermi wavevector onto the plane of rotation. $J_\nu (\mu)$ is the $\nu^\text{th}$ order Bessel function of the first kind. This equation still holds approximately for low levels of Fermi surface warping\cite{Yagi}, as in the current work. The oscillatory nature of the Bessel functions leads to oscillations in the magnetoresistance. In order to link these oscillations to the form of the FS, we make two assumptions: firstly that $\omega_c \tau \cos \theta$ is large enough to neglect all terms in the sum except $J_0(\mu)$, and secondly that $\mu\gg1$ so that we can expand $J_0(\mu)$, in order to obtain expressions for the zeros of $\frac{\sigma_{zz}(\theta)}{\sigma_{zz}(0)}$. These zeros are at $\mu = n \pi + \frac{\pi}{2} \pm \frac{\pi}{4}$, where the positive sign is for $\mu<0$ and the negative for $\mu>0$, known as the Yamaji angles \cite{Yamaji,Lewin}, allowing us to directly link the positions of AMRO peaks to the form of the FS. The first approximation will break down for $\theta$ approaching $90^\circ$, whilst the second will break down if $\theta$ becomes too small.

\section{Results}

\begin{figure}[htbp]
\centering
  \includegraphics[width=0.45\textwidth]{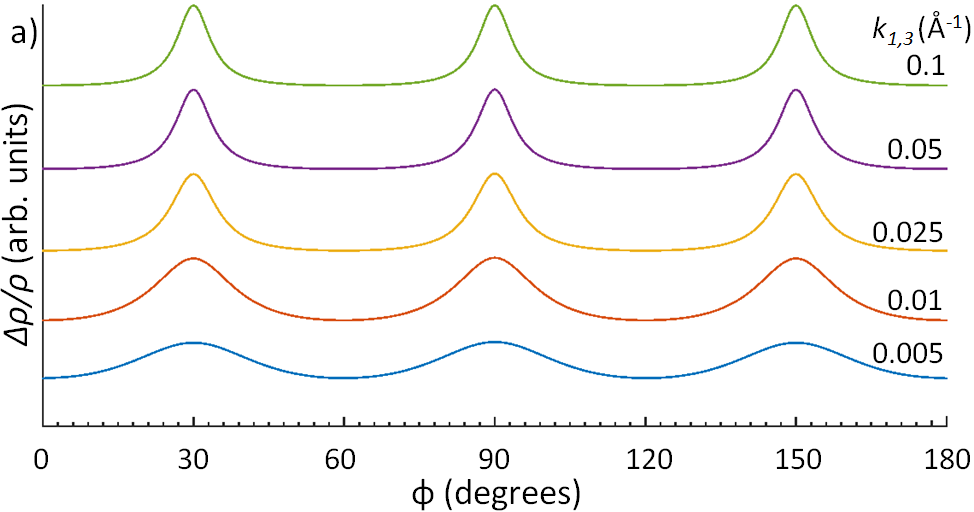}
   \includegraphics[width=0.45\textwidth]{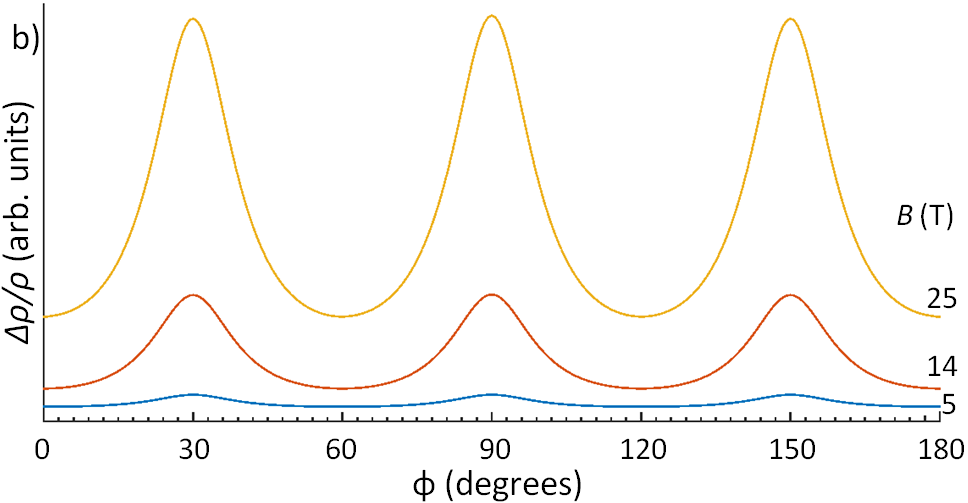}
   \includegraphics[width=0.45\textwidth]{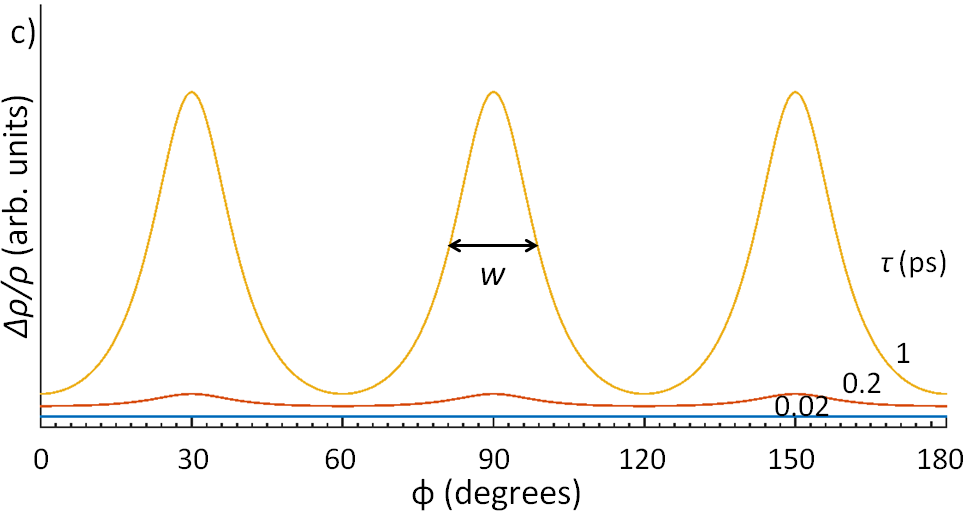}
   \includegraphics[width=0.45\textwidth]{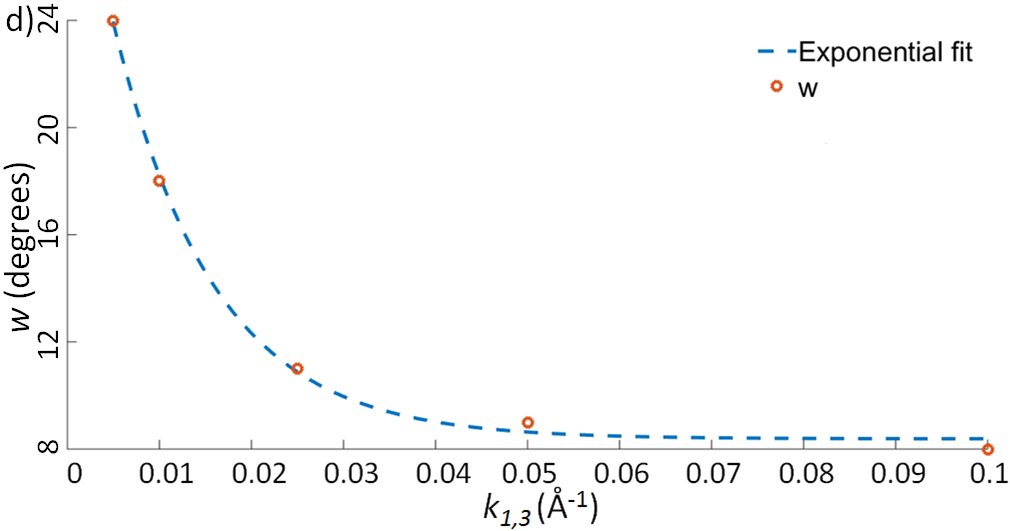}
\caption{(colour online) Simulated AMROs as a function of azimuthal angle, $\phi$,
for different values of  a) the $c$-axis warping $k_{1,3}=0.005-0.1$~$\text{\r{A}}^{-1}$, b) the magnetic field strength, $B=5-25$~T, and c) the isotropic scattering time, $\tau=0.02-1$~ps.
Each set of AMRO is normalised by the range of $\frac{\Delta \rho}{\rho}$ values within the set and shifted vertically for clarity,
and the constant parameters used in simulations
 were $k_{1,3}=0.01$~$\text{\r{A}}^{-1}$, $\tau=10$~ps, $m^*=1.5m_e$ and $B=25$~T.
  The exponential effect of warping $k_{1,3}$ on the full width at half maximum of the peaks, $w$, (defined in c)
  is shown in d). Dashed line is an exponential fit of the
  form $w=a e^{-bk_{1,3}} +c$ with $a=24.493^\circ$, $b=91.346$~\r{A}, and $c=8.3818^\circ$.
} \label{fig:InPlaneVaryingk31}
\end{figure}

\subsection{The angular magnetoresistance oscillations due to in-plane rotation}

We now present the results of our AMRO simulations obtained when the magnetic field is rotated in the ($ab$)-plane as a function of the azimuthal angle, $\phi$, for different parameters, as shown in Fig. \ref{fig:InPlaneVaryingk31}. First of all, the AMRO spectra have a $60^\circ$ periodicity in $\phi$, as would be expected from the hexagonal symmetry, and the AMRO peaks become sharper and more prominent as the degree of the inter-plane $c$-axis warping, $k_{1,3}$, increases (see Fig. \ref{fig:InPlaneVaryingk31}a). This variation can be quantified by the variation of its full width at half maximum $w$, as a function of  $k_{1,3}$ which clearly shows an exponential decay as the peaks become sharper and sharper (see Fig. \ref{fig:InPlaneVaryingk31}d). This could potentially be caused by how the integral for $\sigma_{zz}$ contains a factor of $\frac{1}{k_F}$ - for low levels of $c$-axis warping, the value of $k_F$ decreases almost linearly as we move away from the corners of the hexagonal prism, whilst for high levels of warping, $k_F$ decreases more quickly as we move away from the corners. This means that $\frac{\Delta \rho}{\rho}\propto\frac{1}{\sigma_{zz}}$ will behave similarly, leading to the width of the peaks reducing as warping increases.

Next, we investigate the effect of the magnetic field strength, $B$, and isotropic scattering time, $\tau$,
on the in-plane AMRO as a function of the azimuthal angle, $\phi$. We observe a continuous enhancement
of magnetoresistance as a function of both $B$ and $\tau$, as shown in Fig. \ref{fig:InPlaneVaryingk31}b) and c).  The quantity $\omega_c \tau = \frac{eB\tau}{m^*}$ dictates how far a quasiparticle travels before scattering, as can be seen in the exponential within the integrand of Eq. \eqref{eq:SCTI}, and thus controls its contribution to the integral giving $\sigma_{zz}$. At the minima of the in-plane AMRO, $v_z$ is essentially zero, making $\sigma_{zz}$ independent of $\omega_c \tau$, but at the maxima the integral in Eq. \eqref{eq:SCTI} of the sinusoidal $v_z$ and the exponential factor means $\sigma_{zz}$ decreases with increasing $\omega_c \tau$ for $\omega_c \tau \gg 1$ (see Appendix D). This means the AMRO peaks increase with $B$ and $\tau$, as seen. Magnetoresistance oscillations are often plotted against $\omega_c \tau$ itself \cite{Lewin,Yakovenko}; in this work, the dependences on the magnetic field and the scattering time are considered separately, to show that both parameters have an effect and to enable direct comparison with previous work in which these parameters are considered individually \cite{Takatsu,Kikugawa}.
The variation of the simulated AMRO with the strength of the magnetic field is similar to that measured experimentally on PdCoO$_2$ in Ref. \onlinecite{Takatsu}, and also strongly resemble those calculated in Ref. \onlinecite{Takatsu}, which have an almost sinusoidal curve with a $60^\circ$ periodicity in $\phi$. The best correspondence between the AMRO of this work and Ref. \onlinecite{Takatsu} is for low values of $k_{1,3}=0.005-0.01$~$\text{\r{A}}^{-1}$, somewhat larger than the experimental value of $0.001$~$\text{\r{A}}^{-1}$ extracted from quantum oscillations\cite{Hicks}.

\subsection{The angular magnetoresistance oscillations due to out-of-plane rotation}

\begin{figure}[htbp]
\centering
\includegraphics[width=0.5\textwidth]{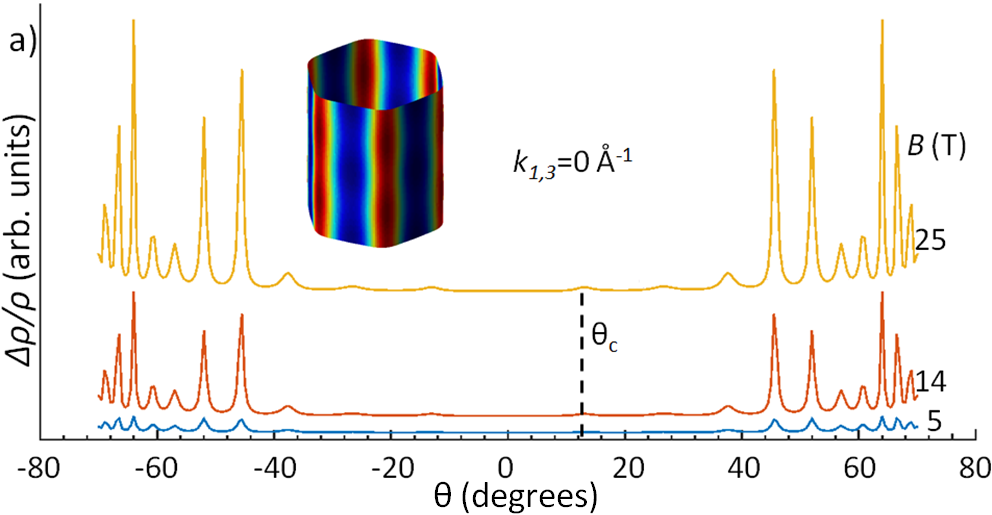}
\includegraphics[width=0.5\textwidth]{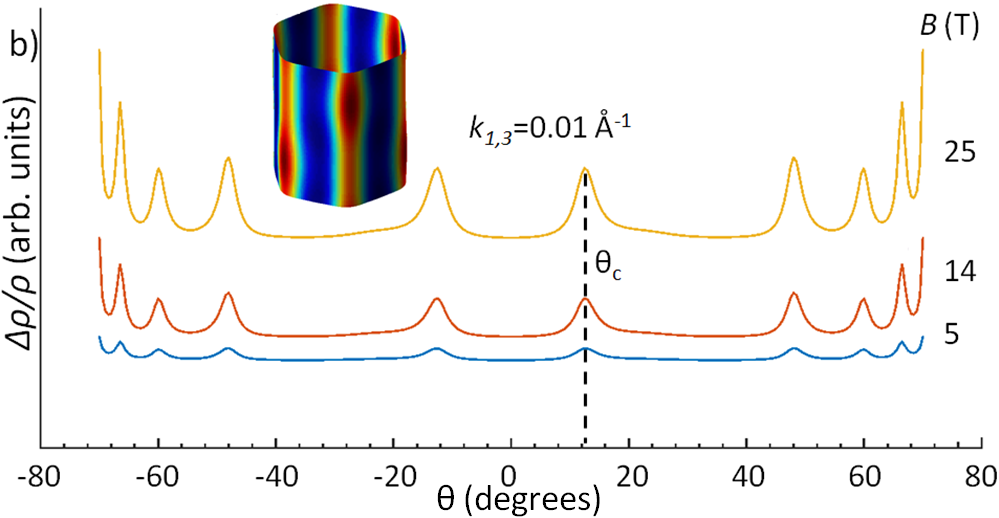}
\includegraphics[width=0.5\textwidth]{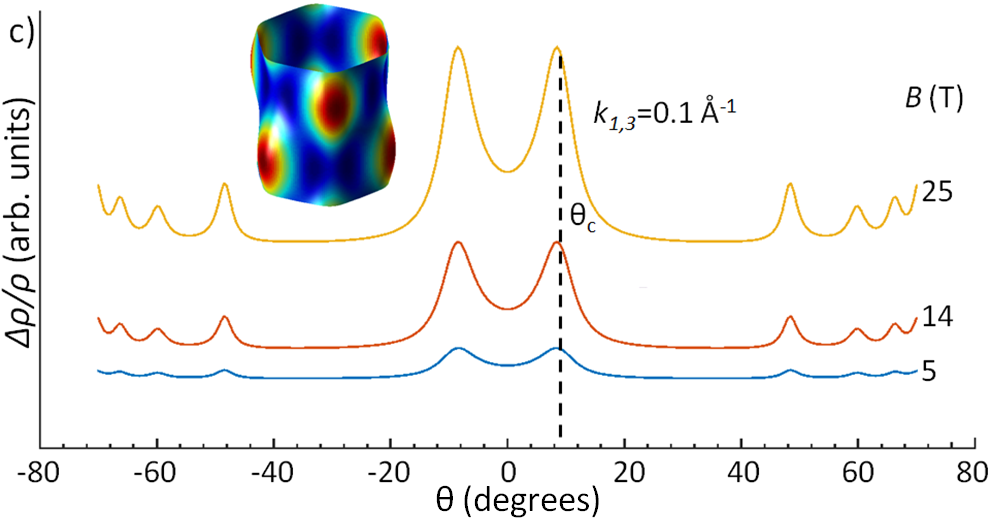}
\includegraphics[width=0.5\textwidth]{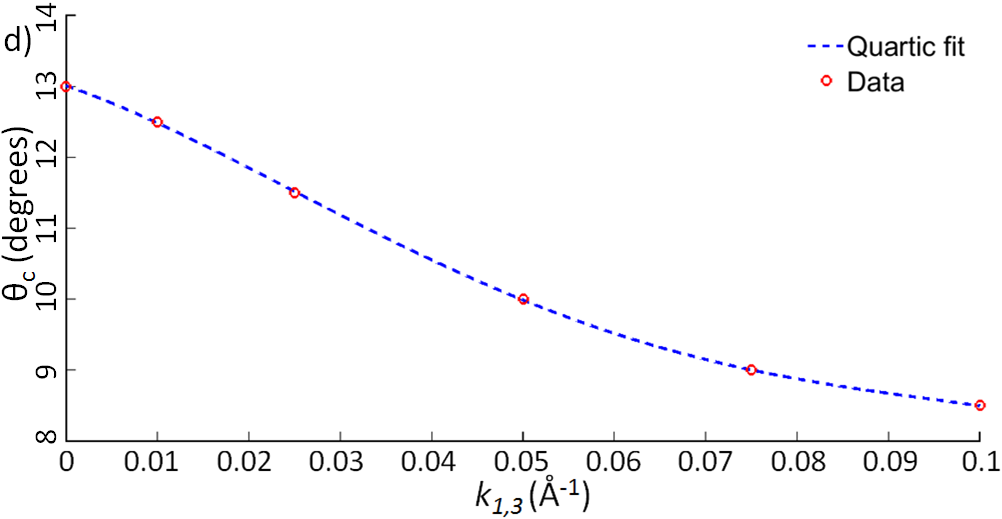}
\caption{(colour online)  Simulated AMROs as a function of polar angle, $\theta$,
for at fixed magnetic field strengths, $B$=5, 14, 25~T,
and different values of the degree of $c$-axis warping, $k_{1,3}$, as follows:
a) 0, b) 0.01 and c) 0.1~$\text{\r{A}}^{-1}$, respectively
and $\phi=0^\circ$, $\tau=10$ps and $m^*=1.5m_e$.
Insets show the corresponding FS coloured by Fermi velocity, given by Eq. \eqref{eq:kfexpansion}
and the vertical dashed line indicates the position of the dominant
central peak at an angle $\theta_c$.
The variation of $\theta_c$ with $k_{1,3}$, is shown in d).
The dotted line is a quartic fit of the form $ak_{1,3}^4+bk_{1,3}^3+ck_{1,3}^2+dk_{1,3}+e$,
with $a=-69302$~$\text{\r{A}}^4$, $b=16639$~$\text{\r{A}}^3$, $c=-977.66$~$\text{\r{A}}^2$, $d=-44.406$~\r{A} and $e=13.007$.}
\label{fig:Varyingk31andB}
\end{figure}

The most common way of using AMRO to extract information about the Fermi surface shape is looking at the effect as a function of the rotation of the magnetic field out of the ($ab$)-plane, from $\theta=0$ ($B \parallel c$) to close to $\theta=90^\circ$ ($B \perp c$), as a function of the magnetic field strength, $B$, and the degree of $c$-axis warping, $k_{1,3}$, as shown in Fig. \ref{fig:Varyingk31andB}. We observe that the magnitude of the AMRO peaks increases with increasing magnetic field, as would be expected from Eq. \eqref{eq:McKenzieForm},  as the quasiparticle are able to move further around their orbits before scattering, as mentioned previously. For a cylindrical Fermi surface with no warping, the peaks in magnetoresistance have the largest amplitude at higher $\theta$ angles (see Fig. \ref{fig:Varyingk31andB}a), as observed for experiments and calculations in other systems, due to the contribution of closed orbits to the magnetoresistance \cite{Goddard,Hussey,Analytis}. When the $c$-axis warping, $k_{1,3}$, increases, we observe that the peaks closest to $\theta=0^\circ$ begin to appear and to grow, dominating over the standard AMRO peaks visible at higher angles. These features show similarities to those observed experimentally for hexagonal PdCoO$_2$ and PtCoO$_2$\cite{Kikugawa,Kushwaha}, although they are at lower values of $\theta$ than the AMRO observed in other materials \cite{Goddard}, where the approximation $\mu\gg1$ will no longer hold. As the warping increases, the central peaks move toward $\theta=0$ (for $\phi=0$) whilst these other peaks move outwards, towards even higher $\theta$, as shown in Fig. \ref{fig:Varyingk31andB}a)-c). The movement of these peaks can be explained  by considering Yamaji angles \cite{Yamaji} - certain values of $\theta$ at which all quasiparticle orbits have the same area. It can be shown that the average value of $v_z$ around a quasiparticle orbit is proportional to the derivative of the orbit's area with respect to the $k_z$ position of the orbit\cite{Singleton}. For Yamaji angles the orbit area is a constant, so $v_z$ averages to zero, giving a minimum in $\sigma_{zz}$ and thus a maximum in $\rho_{zz}$\cite{Bergemann,Yagi}. As the shape of the Fermi surface changes, so too do the Yamaji angles at which the AMRO peaks occur, as seen. The position of the central peaks, $\theta_c$, changes slightly as the $c$-axis warping changes, and it seems to start to plateau as $k_{1,3}$ becomes large (see Fig. \ref{fig:InPlaneVaryingk31}d). This decay is fitted to a quartic function, which could then be used to estimate the position of the central peaks for a given $c$-axis warping and $\phi$.

\begin{figure}[h!]
\centering
\includegraphics[width=0.47\textwidth]{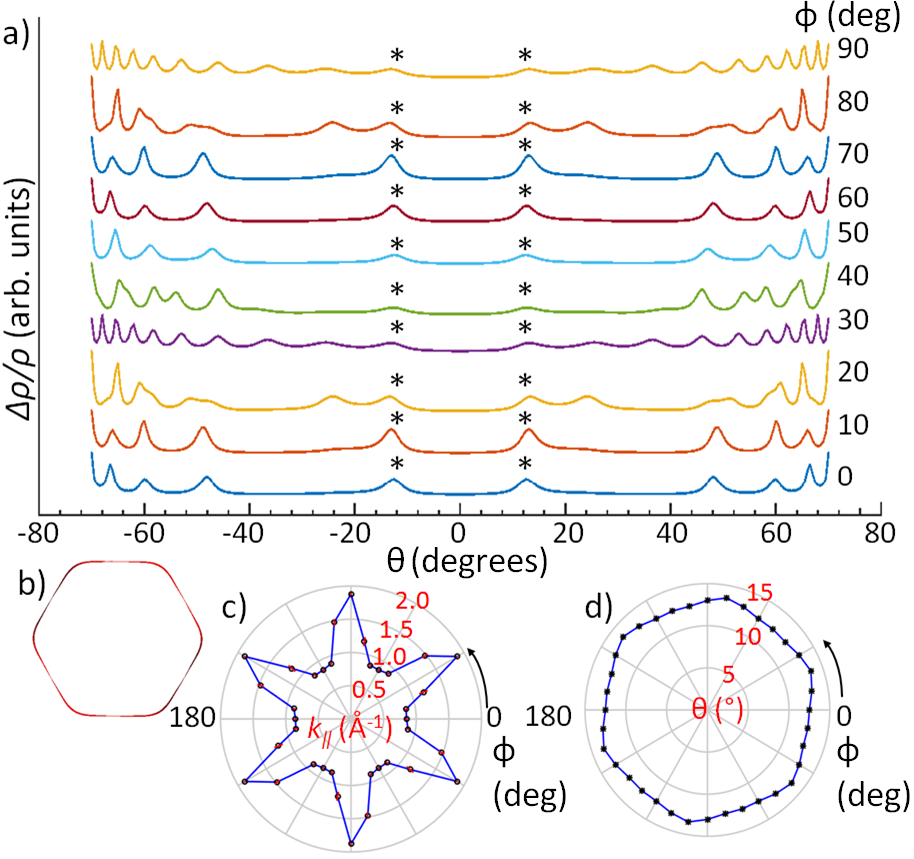}
\includegraphics[width=0.47\textwidth]{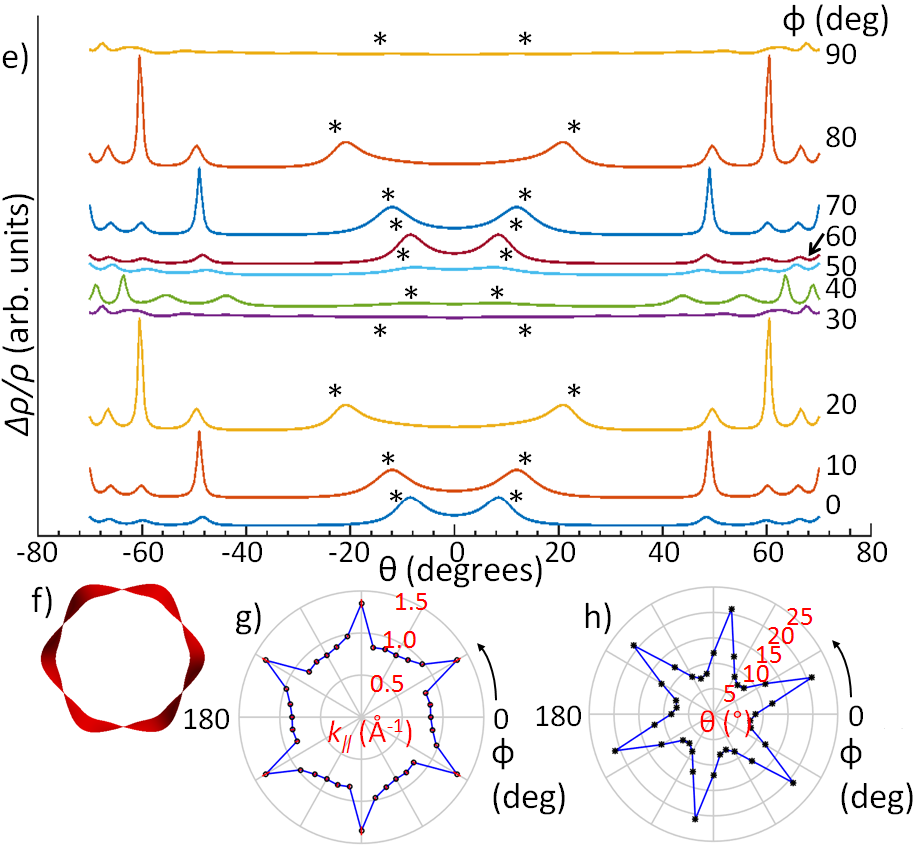}
\caption{(colour online) The simulated AMROs as a function of $\theta$ for given values of the azimuthal angle $\phi$, and two
different $c$-axis warping a) $k_{1,3}=0.01\text{\r{A}}^{-1}$ and e) $0.1\text{\r{A}}^{-1}$ corresponding to the
 Fermi surfaces shown as a top-down views in b) and f), respectively and using  $B=25$T, $\tau=10$ps and $m^*=1.5m_e$.  The polar plots of $k_\parallel$, against azimuthal angle $\phi$ (extracted as detailed in Appendix G) is shown in c) for $k_{1,3}=0.01$ and g) for $k_{1,3}=0.1\text{\r{A}}^{-1}$, respectively and the error bars are shown in red. The position of the central peaks, $\theta_c$, is marked with asterisks and its variation with $\phi$ for each value of $k_{1,3}$ is shown separately in d) and h), respectively.}
   \label{fig:VaryingPhi}
\end{figure}

We extend the above AMRO simulations to include the effect of changing the azimuthal angle of the magnetic field direction, $\phi$, which is necessary to explore the shape of the Fermi surface.
Fig.~\ref{fig:VaryingPhi}a) and d) show the simulated AMRO as a function of $\theta$ for various values of $\phi$ at fixed magnetic field $B=25$~T and for two different values of
$k_{1,3}=0.01$ and $0.1$~$\text{\r{A}}^{-1}$, respectively.
A clear $60^\circ$ periodicity in $\phi$ can be observed in both graphs by comparing the results from $0^\circ$ up to $30^\circ$ to those from $60^\circ$ up to $90^\circ$, as would be expected due to the periodicity of the FS itself.
The amplitude of the AMRO peaks, especially the central ones,
 as a function of $\phi$ is strongly sensitive to the exact value of $k_{1,3}$, being strongest near $\phi=0^\circ$, and weakest near $\phi=30^\circ$ in Fig. \ref{fig:VaryingPhi} (which are very similar to those for no warping in Fig.~\ref{fig:Varyingk31andB}a). Based on Eq. \eqref{eq:kfexpansion}, the $c$-axis warping reaches its maximum and minimum magnitude respectively at these $\phi$ points, which results in the AMROs being maximally and minimally different from the zero warping case.
At high $\theta$ values approaching $\theta=90^\circ$, the sharp AMROs become more tightly spaced and generally larger in amplitude as $k_{1,3}$ decreases (although there are some peaks that actually decrease in magnitude), and also strongly change their positions
 as a function of $\phi$, especially for higher warping, as shown in Fig.~\ref{fig:VaryingPhi}e).

Extracting the shape of the Fermi surface from angle dependent magnetoresistance oscillation data is done by considering the values of $\theta$ at the maximum peaks in resistivity. It can be shown that as the magnetic field is rotated from the $c$-axis to the ($ab$)-plane, the position of peaks in the magnetoresistance can be linked to $k_\parallel$, the \emph{maximum} projection of a Fermi wavevector onto the plane defined by the rotation of $\mathbf{B}$\cite{McKenzie, Singleton,Yamaji}. This allows us to extract the shape of the FS as we rotate the direction of the magnetic field, and thus the direction of the projected vector, in the ($ab$)-plane \cite{Singleton}.
This is the core method behind using AMRO to map out Fermi surfaces,
which is described in detail in the Appendices.

Fig.~\ref{fig:VaryingPhi}b) and e) show the variation with $\phi$ of the maximum possible projection of an in-plane Fermi wavevector onto the plane of rotation, $k_\parallel$, as extracted by fitting simulated data to the Yamaji angle formula (Eq. \eqref{eq:kparallelequation} in Appendices), for $k_{1,3}$=0.01 and 0.1$\text{\r{A}}^{-1}$, respectively. In general, when $\tan \theta_n$ is plotted against $n \pm \frac{1}{4}$, the points fit well to a straight line, giving a reliable value of $k_\parallel$, as shown in Fig. \ref{fig:KParFitting} in the Appendices. The central peaks are almost always just off this line, as would be expected due to the breakdown of the approximation that $\mu\gg1$, meaning that the formula $\mu = n\pi+\frac{\pi}{2}\pm\frac{\pi}{4}$ is no longer a good fit to the positions of the peaks. In the case of $30^\circ , 90^\circ, $ etc., the central peaks have disappeared. If the central peaks are included when fitting for $k_\parallel$, the error in the fitted gradient increases by an order of magnitude, but the fit is indistinguishable from the fit without the central peaks included.
The polar plots of the  $k_\parallel$ shows a $60^\circ$ periodicity in $\phi$, as expected from the periodicity in AMROs and the lattice in Fig.~\ref{fig:VaryingPhi}b) and e) and these shapes are different from the
in-plane shape of the hexagonal Fermi surface, as the locus of $k_\parallel$ does not have to match the outline of the FS, as seen previously for organic conducting salts \cite{Singleton,Goddard}.
Furthermore, the variation of the position of the central peaks, $\theta_c$, with $\phi$, shown in Fig. \ref{fig:VaryingPhi}c) and f)
 also has a $60^\circ$ periodicity and for small warping is quite similar to the in-plane shape of the hexagonal Fermi surface. As the degree of warping increases, the range of $\theta_c$ values becomes much larger, shifting from around $1^\circ$ for $k_{1,3}=0.01$~$\text{\r{A}}^{-1}$ to around $15^\circ$ for $k_{1,3}=0.1$~$\text{\r{A}}^{-1}$.

 Even for simple Fermi surfaces, the locus of $k_\parallel$ may have a complicated shape, which makes it difficult to find an appropriate functional form. In the case of a simple ellipse \cite{Singleton}, the locus is a more complicated dumbbell-like shape, and for more complex Fermi surfaces, we may expect even more complex loci for $k_\parallel$. Using this to trace out the shape of the Fermi surface is not a trivial task, becoming more difficult for higher degrees of warping, as is the case in the simulations presented here.
These simulations of the polar plots of $k_\parallel$ together with those of the raw AMRO data in Figs. \ref{fig:Varyingk31andB} and \ref{fig:VaryingPhi}, can be compared to future experimental results to identify the relevant parameters related to the Fermi surface in layered hexagonal materials.

\begin{figure}[t]
\centering
\includegraphics[width=0.5\textwidth]{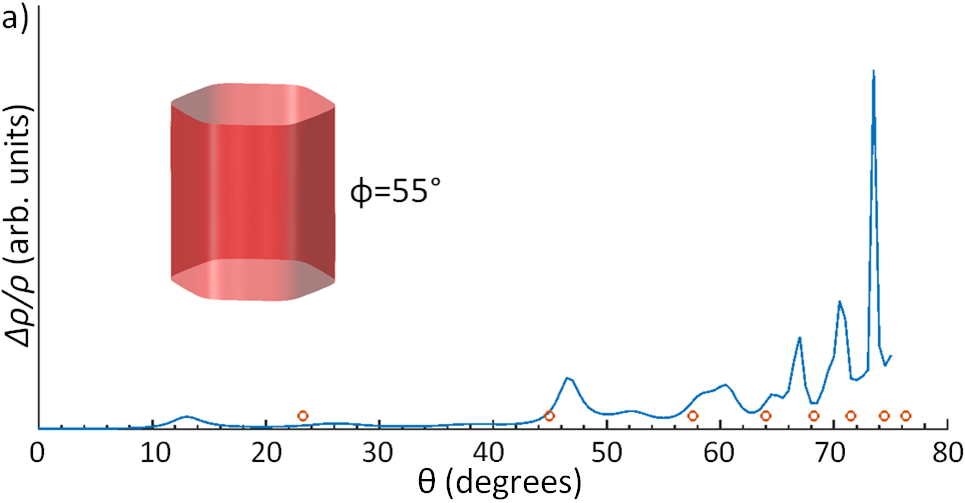}
\includegraphics[width=0.5\textwidth]{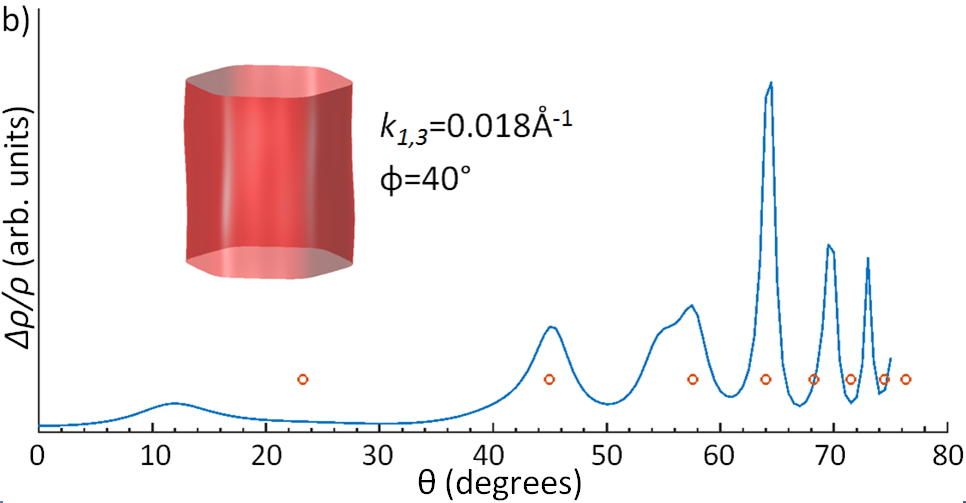}
\caption{(colour online) a) Simulated AMROs for a realistic Fermi surface of PdCoO$_2$
using $\phi=55^{\circ}$ and $k_{00}=0.9559, k_{0,6}=0.04, k_{0,12}=0.007$ and $k_{2,0}=-0.0025$
and $k_{1,3}=0.001$~(in units of $\text{\r{A}}^{-1}$), as determined from quantum oscillations \cite{Hicks}.
b) Simulated AMROs for $k_{1,3}=0.018$~$\text{\r{A}}^{-1}$, $\phi=40^\circ$ to best match the positions of the AMRO peaks measured for PdCoO$_2$ in Ref.\onlinecite{Kikugawa}. The positions of the experimental AMRO peaks are marked by red circles
and the other parameters used in simulations were $B=25$T, $\tau=10$ps and $m^*=1.5m_e$. A complete experimental angular
dependence in ($\theta$, $\phi$) is necessary to constrain the large range of parameters. } \label{fig:ExperimentalFit}
\end{figure}

Finally, we simulate AMROs for the Fermi surface of PdCo$_2$, calculated based on quantum oscillation data \cite{Hicks} (although we do not including the $k_{0,1}$ term), as shown in Fig. \ref{fig:ExperimentalFit}a), with $\phi$ taken as $55^\circ$ in order to best match the positions of AMRO peaks observed experimentally in Ref.\onlinecite{Kikugawa}.
In order to find a better description for the available experimental data,
we can vary both $k_{1,3}$ and $\phi$ and calculate the AMRO for each set of values,
while minimizing the sum of the squares of the differences between the calculated and experimental peak positions. We find
the best match for $k_{1,3}=0.018$~$\text{\r{A}}^{-1}$ and $\phi=40^\circ$, shown in Fig. \ref{fig:ExperimentalFit}b).
This estimate for $k_{1,3}$ is significantly larger than that measured through quantum oscillations\cite{Hicks}, and slightly outside the range estimated using the in-plane AMRO.  In order to constrain the large number of available parameters in simulations and to better match simulation with experiment, a complete experimental data set is needed.
Furthermore, the effects neglected in our calculation, such as anisotropy in the scattering time $\tau$ or the effective mass $m^*$, may also need to be taken into account in future work, and to obtain the sharp AMRO calculated here, the scattering time needs to be $\tau \gtrsim 0.5$~ps, giving a rough guide to the quality of single crystals required for these studies.
 Despite such considerations, we would expect that certain features in our calculations will be robust measures of Fermi surface topography. In particular the central AMRO peaks are a very robust measure of interplane warping which may be little affected by the inclusion of anisotropy in $\tau$ and $m^*$  \cite{Kennett}. Following their behaviour as a function of $\phi$ would allow the level of warping and the Fermi surface topography of materials with hexagonal lattices to be identified.

\section{Summary}

In summary, we have investigated the angle dependent magnetoresistance oscillations
 corresponding to Fermi surfaces of structures with hexagonal lattices
using a tight-binding expansion of the Fermi surface in terms of cylindrical harmonics.
 We have investigated the effect of varying degrees of $c$-axis warping
 as a function both of in-plane and out-of plane rotations in magnetic field and found
 that the form of the AMRO is very sensitive to the degree of the $c$-axis warping.
 We also find that the dependence of the out-of-plane AMRO peak positions on the
azimuthal angle $\phi$ becomes stronger for a more warped Fermi surface.
We also attempt to compare our simulations to experimental work
on a candidate compound, PdCoO$_2$ \cite{Takatsu,Kikugawa}.
A further extension of this work could include the effect of anisotropy in the scattering time and effective mass. In general these quantities will depend on the position of the particle in reciprocal space, but their functional form must obey the symmetries of the Fermi surface. Finally, these simulations will be of great use to compare with future comprehensive experimental work on  PdCoO$_2$, PtCoO$_2$ and similar compounds, in order to fully extract the topology of their Fermi surfaces from angle dependent magnetoresistance oscillations.

\section{Acknowledgements}
We thank Stephen Blundell for detailed and useful comments on our manuscript.
We would like to acknowledge the use of the University of Oxford Advanced Research Computing (ARC) facility in carrying out this work. This work was mainly supported by EPSRC (EP/L001772/1, EP/I004475/1, EP/I017836/1).
 AIC acknowledges an EPSRC Career Acceleration Fellowship (EP/I004475/1).

\bibliography{}{}

\begin{thebibliography}{34}%
\makeatletter
\providecommand \@ifxundefined [1]{%
 \@ifx{#1\undefined}
}%
\providecommand \@ifnum [1]{%
 \ifnum #1\expandafter \@firstoftwo
 \else \expandafter \@secondoftwo
 \fi
}%
\providecommand \@ifx [1]{%
 \ifx #1\expandafter \@firstoftwo
 \else \expandafter \@secondoftwo
 \fi
}%
\providecommand \natexlab [1]{#1}%
\providecommand \enquote  [1]{``#1''}%
\providecommand \bibnamefont  [1]{#1}%
\providecommand \bibfnamefont [1]{#1}%
\providecommand \citenamefont [1]{#1}%
\providecommand \href@noop [0]{\@secondoftwo}%
\providecommand \href [0]{\begingroup \@sanitize@url \@href}%
\providecommand \@href[1]{\@@startlink{#1}\@@href}%
\providecommand \@@href[1]{\endgroup#1\@@endlink}%
\providecommand \@sanitize@url [0]{\catcode `\\12\catcode `\$12\catcode
  `\&12\catcode `\#12\catcode `\^12\catcode `\_12\catcode `\%12\relax}%
\providecommand \@@startlink[1]{}%
\providecommand \@@endlink[0]{}%
\providecommand \url  [0]{\begingroup\@sanitize@url \@url }%
\providecommand \@url [1]{\endgroup\@href {#1}{\urlprefix }}%
\providecommand \urlprefix  [0]{URL }%
\providecommand \Eprint [0]{\href }%
\providecommand \doibase [0]{http://dx.doi.org/}%
\providecommand \selectlanguage [0]{\@gobble}%
\providecommand \bibinfo  [0]{\@secondoftwo}%
\providecommand \bibfield  [0]{\@secondoftwo}%
\providecommand \translation [1]{[#1]}%
\providecommand \BibitemOpen [0]{}%
\providecommand \bibitemStop [0]{}%
\providecommand \bibitemNoStop [0]{.\EOS\space}%
\providecommand \EOS [0]{\spacefactor3000\relax}%
\providecommand \BibitemShut  [1]{\csname bibitem#1\endcsname}%
\let\auto@bib@innerbib\@empty
\bibitem [{\citenamefont {Damascelli}(2004)}]{Damascelli}%
  \BibitemOpen
  \bibfield  {author} {\bibinfo {author} {\bibfnamefont {A.}~\bibnamefont
  {Damascelli}},\ }\bibfield  {title} {\enquote {\bibinfo {title} {Probing the
  electronic structure of complex systems by {ARPES}},}\ }\href@noop {}
  {\bibfield  {journal} {\bibinfo  {journal} {Physica Scripta}\ }\textbf
  {\bibinfo {volume} {2004}},\ \bibinfo {pages} {61} (\bibinfo {year}
  {2004})}\BibitemShut {NoStop}%
\bibitem [{\citenamefont {Ashcroft}\ and\ \citenamefont
  {Mermin}(1976)}]{Ashcroft}%
  \BibitemOpen
  \bibfield  {author} {\bibinfo {author} {\bibfnamefont {N.~W.}\ \bibnamefont
  {Ashcroft}}\ and\ \bibinfo {author} {\bibfnamefont {N.~D.}\ \bibnamefont
  {Mermin}},\ }\href@noop {} {\emph {\bibinfo {title} {Solid State Physics}}}\
  (\bibinfo  {publisher} {Cengage Learning},\ \bibinfo {address} {Belmont},\
  \bibinfo {year} {1976})\BibitemShut {NoStop}%
\bibitem [{\citenamefont {Goddard}\ \emph {et~al.}(2004)\citenamefont
  {Goddard}, \citenamefont {Blundell}, \citenamefont {Singleton}, \citenamefont
  {McDonald}, \citenamefont {Ardavan}, \citenamefont {Narduzzo}, \citenamefont
  {Schlueter}, \citenamefont {Kini},\ and\ \citenamefont {Sasaki}}]{Goddard}%
  \BibitemOpen
  \bibfield  {author} {\bibinfo {author} {\bibfnamefont {P.~A.}\ \bibnamefont
  {Goddard}}, \bibinfo {author} {\bibfnamefont {S.~J.}\ \bibnamefont
  {Blundell}}, \bibinfo {author} {\bibfnamefont {J.}~\bibnamefont {Singleton}},
  \bibinfo {author} {\bibfnamefont {R.~D.}\ \bibnamefont {McDonald}}, \bibinfo
  {author} {\bibfnamefont {A.}~\bibnamefont {Ardavan}}, \bibinfo {author}
  {\bibfnamefont {A.}~\bibnamefont {Narduzzo}}, \bibinfo {author}
  {\bibfnamefont {J.~A.}\ \bibnamefont {Schlueter}}, \bibinfo {author}
  {\bibfnamefont {A.~M.}\ \bibnamefont {Kini}}, \ and\ \bibinfo {author}
  {\bibfnamefont {T.}~\bibnamefont {Sasaki}},\ }\bibfield  {title} {\enquote
  {\bibinfo {title} {Angle-dependent magnetoresistance of the layered organic
  superconductor
  {$\kappa{}-(\mathrm{ET}{)}_{2}\mathrm{Cu}(\mathrm{NCS}{)}_{2}:$} simulation
  and experiment},}\ }\href@noop {} {\bibfield  {journal} {\bibinfo  {journal}
  {Phys. Rev. B}\ }\textbf {\bibinfo {volume} {69}},\ \bibinfo {pages} {174509}
  (\bibinfo {year} {2004})}\BibitemShut {NoStop}%
\bibitem [{\citenamefont {Singleton}(2000)}]{Singleton}%
  \BibitemOpen
  \bibfield  {author} {\bibinfo {author} {\bibfnamefont {J.}~\bibnamefont
  {Singleton}},\ }\bibfield  {title} {\enquote {\bibinfo {title} {Studies of
  quasi-two-dimensional organic conductors based on {BEDT}-{TTF} using high
  magnetic fields},}\ }\href@noop {} {\bibfield  {journal} {\bibinfo  {journal}
  {Reports on Progress in Physics}\ }\textbf {\bibinfo {volume} {63}},\
  \bibinfo {pages} {1111} (\bibinfo {year} {2000})}\BibitemShut {NoStop}%
\bibitem [{\citenamefont {Enomoto}\ \emph {et~al.}(2006)\citenamefont
  {Enomoto}, \citenamefont {Uji}, \citenamefont {Yamaguchi}, \citenamefont
  {Terashima}, \citenamefont {Konoike}, \citenamefont {Nishimura},
  \citenamefont {Enoki}, \citenamefont {Suzuki},\ and\ \citenamefont
  {Suzuki}}]{Enomoto}%
  \BibitemOpen
  \bibfield  {author} {\bibinfo {author} {\bibfnamefont {K.}~\bibnamefont
  {Enomoto}}, \bibinfo {author} {\bibfnamefont {S.}~\bibnamefont {Uji}},
  \bibinfo {author} {\bibfnamefont {T.}~\bibnamefont {Yamaguchi}}, \bibinfo
  {author} {\bibfnamefont {T.}~\bibnamefont {Terashima}}, \bibinfo {author}
  {\bibfnamefont {T.}~\bibnamefont {Konoike}}, \bibinfo {author} {\bibfnamefont
  {M.}~\bibnamefont {Nishimura}}, \bibinfo {author} {\bibfnamefont
  {T.}~\bibnamefont {Enoki}}, \bibinfo {author} {\bibfnamefont
  {M.}~\bibnamefont {Suzuki}}, \ and\ \bibinfo {author} {\bibfnamefont {I.~S.}\
  \bibnamefont {Suzuki}},\ }\bibfield  {title} {\enquote {\bibinfo {title}
  {Fermi surface and interlayer transport in high-stage
  {$\mathrm{Mo}{\mathrm{Cl}}_{5}$} graphite intercalation compounds},}\
  }\href@noop {} {\bibfield  {journal} {\bibinfo  {journal} {Phys. Rev. B}\
  }\textbf {\bibinfo {volume} {73}},\ \bibinfo {pages} {045115} (\bibinfo
  {year} {2006})}\BibitemShut {NoStop}%
\bibitem [{\citenamefont {Bergemann}\ \emph {et~al.}(2003)\citenamefont
  {Bergemann}, \citenamefont {Mackenzie}, \citenamefont {Julian}, \citenamefont
  {Forsythe},\ and\ \citenamefont {Ohmichi}}]{Bergemann}%
  \BibitemOpen
  \bibfield  {author} {\bibinfo {author} {\bibfnamefont {C.}~\bibnamefont
  {Bergemann}}, \bibinfo {author} {\bibfnamefont {A.~P.}\ \bibnamefont
  {Mackenzie}}, \bibinfo {author} {\bibfnamefont {S.~R.}\ \bibnamefont
  {Julian}}, \bibinfo {author} {\bibfnamefont {D.}~\bibnamefont {Forsythe}}, \
  and\ \bibinfo {author} {\bibfnamefont {E.}~\bibnamefont {Ohmichi}},\
  }\bibfield  {title} {\enquote {\bibinfo {title} {Quasi-two-dimensional
  {Fermi} liquid properties of the unconventional superconductor
  {$\mathrm{Sr}_{2}\mathrm{RuO}_{4}$}},}\ }\href@noop {} {\bibfield  {journal}
  {\bibinfo  {journal} {Adv. Phys.}\ }\textbf {\bibinfo {volume} {52}},\
  \bibinfo {pages} {639} (\bibinfo {year} {2003})}\BibitemShut {NoStop}%
\bibitem [{\citenamefont {Kimata}\ \emph {et~al.}(2010)\citenamefont {Kimata},
  \citenamefont {Terashima}, \citenamefont {Kurita}, \citenamefont {Satsukawa},
  \citenamefont {Harada}, \citenamefont {Kodama}, \citenamefont {Sato},
  \citenamefont {Imai}, \citenamefont {Kihou}, \citenamefont {Lee},
  \citenamefont {Kito}, \citenamefont {Eisaki}, \citenamefont {Iyo},
  \citenamefont {Saito}, \citenamefont {Fukazawa}, \citenamefont {Kohori},
  \citenamefont {Harima},\ and\ \citenamefont {Uji}}]{Kimata}%
  \BibitemOpen
  \bibfield  {author} {\bibinfo {author} {\bibfnamefont {M.}~\bibnamefont
  {Kimata}}, \bibinfo {author} {\bibfnamefont {T.}~\bibnamefont {Terashima}},
  \bibinfo {author} {\bibfnamefont {N.}~\bibnamefont {Kurita}}, \bibinfo
  {author} {\bibfnamefont {H.}~\bibnamefont {Satsukawa}}, \bibinfo {author}
  {\bibfnamefont {A.}~\bibnamefont {Harada}}, \bibinfo {author} {\bibfnamefont
  {K.}~\bibnamefont {Kodama}}, \bibinfo {author} {\bibfnamefont
  {A.}~\bibnamefont {Sato}}, \bibinfo {author} {\bibfnamefont {M.}~\bibnamefont
  {Imai}}, \bibinfo {author} {\bibfnamefont {K.}~\bibnamefont {Kihou}},
  \bibinfo {author} {\bibfnamefont {C.~H.}\ \bibnamefont {Lee}}, \bibinfo
  {author} {\bibfnamefont {H.}~\bibnamefont {Kito}}, \bibinfo {author}
  {\bibfnamefont {H.}~\bibnamefont {Eisaki}}, \bibinfo {author} {\bibfnamefont
  {A.}~\bibnamefont {Iyo}}, \bibinfo {author} {\bibfnamefont {T.}~\bibnamefont
  {Saito}}, \bibinfo {author} {\bibfnamefont {H.}~\bibnamefont {Fukazawa}},
  \bibinfo {author} {\bibfnamefont {Y.}~\bibnamefont {Kohori}}, \bibinfo
  {author} {\bibfnamefont {H.}~\bibnamefont {Harima}}, \ and\ \bibinfo {author}
  {\bibfnamefont {S.}~\bibnamefont {Uji}},\ }\bibfield  {title} {\enquote
  {\bibinfo {title} {Quasi-two-dimensional {Fermi} surfaces and coherent
  interlayer transport in {${\mathrm{KFe}}_{2}{\mathrm{As}}_{2}$}},}\
  }\href@noop {} {\bibfield  {journal} {\bibinfo  {journal} {Phys. Rev. Lett.}\
  }\textbf {\bibinfo {volume} {105}},\ \bibinfo {pages} {246403} (\bibinfo
  {year} {2010})}\BibitemShut {NoStop}%
\bibitem [{\citenamefont {Abdel-Jawad}\ \emph {et~al.}(2006)\citenamefont
  {Abdel-Jawad}, \citenamefont {Kennett}, \citenamefont {Balicas},
  \citenamefont {Carrington}, \citenamefont {Mackenzie}, \citenamefont
  {McKenzie},\ and\ \citenamefont {Hussey}}]{Abdel-Jawad}%
  \BibitemOpen
  \bibfield  {author} {\bibinfo {author} {\bibfnamefont {M.}~\bibnamefont
  {Abdel-Jawad}}, \bibinfo {author} {\bibfnamefont {M.~P.}\ \bibnamefont
  {Kennett}}, \bibinfo {author} {\bibfnamefont {L.}~\bibnamefont {Balicas}},
  \bibinfo {author} {\bibfnamefont {A.}~\bibnamefont {Carrington}}, \bibinfo
  {author} {\bibfnamefont {A.~P.}\ \bibnamefont {Mackenzie}}, \bibinfo {author}
  {\bibfnamefont {R.~H.}\ \bibnamefont {McKenzie}}, \ and\ \bibinfo {author}
  {\bibfnamefont {N.~E.}\ \bibnamefont {Hussey}},\ }\bibfield  {title}
  {\enquote {\bibinfo {title} {Anisotropic scattering and anomalous
  normal-state transport in a high-temperature superconductor},}\ }\href@noop
  {} {\bibfield  {journal} {\bibinfo  {journal} {Nat. Phys.}\ }\textbf
  {\bibinfo {volume} {2}},\ \bibinfo {pages} {821--825} (\bibinfo {year}
  {2006})}\BibitemShut {NoStop}%
\bibitem [{\citenamefont {Hussey}\ \emph {et~al.}(2003)\citenamefont {Hussey},
  \citenamefont {Abdel-Jawad}, \citenamefont {Carrington}, \citenamefont
  {Mackenzie},\ and\ \citenamefont {Balicas}}]{Hussey}%
  \BibitemOpen
  \bibfield  {author} {\bibinfo {author} {\bibfnamefont {N.~E.}\ \bibnamefont
  {Hussey}}, \bibinfo {author} {\bibfnamefont {M.}~\bibnamefont {Abdel-Jawad}},
  \bibinfo {author} {\bibfnamefont {A.}~\bibnamefont {Carrington}}, \bibinfo
  {author} {\bibfnamefont {A.~P.}\ \bibnamefont {Mackenzie}}, \ and\ \bibinfo
  {author} {\bibfnamefont {L.}~\bibnamefont {Balicas}},\ }\bibfield  {title}
  {\enquote {\bibinfo {title} {A coherent three-dimensional {Fermi} surface in
  a high-transition-temperature superconductor},}\ }\href@noop {} {\bibfield
  {journal} {\bibinfo  {journal} {Nature}\ }\textbf {\bibinfo {volume} {425}},\
  \bibinfo {pages} {814} (\bibinfo {year} {2003})}\BibitemShut {NoStop}%
\bibitem [{\citenamefont {Analytis}\ \emph {et~al.}(2007)\citenamefont
  {Analytis}, \citenamefont {Abdel-Jawad}, \citenamefont {Balicas},
  \citenamefont {French},\ and\ \citenamefont {Hussey}}]{Analytis}%
  \BibitemOpen
  \bibfield  {author} {\bibinfo {author} {\bibfnamefont {J.~G.}\ \bibnamefont
  {Analytis}}, \bibinfo {author} {\bibfnamefont {M.}~\bibnamefont
  {Abdel-Jawad}}, \bibinfo {author} {\bibfnamefont {L.}~\bibnamefont
  {Balicas}}, \bibinfo {author} {\bibfnamefont {M.~M.~J.}\ \bibnamefont
  {French}}, \ and\ \bibinfo {author} {\bibfnamefont {N.~E.}\ \bibnamefont
  {Hussey}},\ }\bibfield  {title} {\enquote {\bibinfo {title} {Angle-dependent
  magnetoresistance measurements in
  {${\mathrm{Tl}}_{2}{\mathrm{Ba}}_{2}\mathrm{Cu}{\mathrm{O}}_{6+\delta{}}$}
  and the need for anisotropic scattering},}\ }\href@noop {} {\bibfield
  {journal} {\bibinfo  {journal} {Phys. Rev. B}\ }\textbf {\bibinfo {volume}
  {76}},\ \bibinfo {pages} {104523} (\bibinfo {year} {2007})}\BibitemShut
  {NoStop}%
\bibitem [{\citenamefont {Kang}\ and\ \citenamefont {Chung}(2009)}]{Kang}%
  \BibitemOpen
  \bibfield  {author} {\bibinfo {author} {\bibfnamefont {W.}~\bibnamefont
  {Kang}}\ and\ \bibinfo {author} {\bibfnamefont {O.-H.}\ \bibnamefont
  {Chung}},\ }\bibfield  {title} {\enquote {\bibinfo {title}
  {Quasi-one-dimensional {Fermi} surface of
  {${(\text{TMTSF})}_{2}{\text{NO}}_{3}$}},}\ }\href@noop {} {\bibfield
  {journal} {\bibinfo  {journal} {Phys. Rev. B}\ }\textbf {\bibinfo {volume}
  {79}},\ \bibinfo {pages} {045115} (\bibinfo {year} {2009})}\BibitemShut
  {NoStop}%
\bibitem [{\citenamefont {Lewin}\ and\ \citenamefont {Analytis}(2015)}]{Lewin}%
  \BibitemOpen
  \bibfield  {author} {\bibinfo {author} {\bibfnamefont {S.~K.}\ \bibnamefont
  {Lewin}}\ and\ \bibinfo {author} {\bibfnamefont {J.~G.}\ \bibnamefont
  {Analytis}},\ }\bibfield  {title} {\enquote {\bibinfo {title}
  {Angle-dependent magnetoresistance oscillations of cuprate superconductors in
  a model with {Fermi} surface reconstruction and magnetic breakdown},}\
  }\href@noop {} {\bibfield  {journal} {\bibinfo  {journal} {Phys. Rev. B}\
  }\textbf {\bibinfo {volume} {92}},\ \bibinfo {pages} {195130} (\bibinfo
  {year} {2015})}\BibitemShut {NoStop}%
\bibitem [{\citenamefont {Nowojewski}\ \emph {et~al.}(2008)\citenamefont
  {Nowojewski}, \citenamefont {Goddard},\ and\ \citenamefont
  {Blundell}}]{Nowojewski_2008}%
  \BibitemOpen
  \bibfield  {author} {\bibinfo {author} {\bibfnamefont {A.}~\bibnamefont
  {Nowojewski}}, \bibinfo {author} {\bibfnamefont {P.~A.}\ \bibnamefont
  {Goddard}}, \ and\ \bibinfo {author} {\bibfnamefont {S.~J.}\ \bibnamefont
  {Blundell}},\ }\bibfield  {title} {\enquote {\bibinfo {title} {Effect of
  magnetic breakdown on angle-dependent magnetoresistance in a
  quasi-two-dimensional metal: {An} analytically solvable model},}\ }\href@noop
  {} {\bibfield  {journal} {\bibinfo  {journal} {Phys. Rev. B}\ }\textbf
  {\bibinfo {volume} {77}},\ \bibinfo {pages} {012402} (\bibinfo {year}
  {2008})}\BibitemShut {NoStop}%
\bibitem [{\citenamefont {Nowojewski}\ and\ \citenamefont
  {Blundell}(2010)}]{Nowojewski_2010}%
  \BibitemOpen
  \bibfield  {author} {\bibinfo {author} {\bibfnamefont {A.}~\bibnamefont
  {Nowojewski}}\ and\ \bibinfo {author} {\bibfnamefont {S.~J.}\ \bibnamefont
  {Blundell}},\ }\bibfield  {title} {\enquote {\bibinfo {title} {Analytical
  treatment of in-plane magnetotransport in the {Falicov}-{Sievert} model},}\
  }\href@noop {} {\bibfield  {journal} {\bibinfo  {journal} {Phys. Rev. B}\
  }\textbf {\bibinfo {volume} {82}},\ \bibinfo {pages} {075121} (\bibinfo
  {year} {2010})}\BibitemShut {NoStop}%
\bibitem [{\citenamefont {Takatsu}\ \emph {et~al.}(2013)\citenamefont
  {Takatsu}, \citenamefont {Ishikawa}, \citenamefont {Yonezawa}, \citenamefont
  {Yoshino}, \citenamefont {Shishidou}, \citenamefont {Oguchi}, \citenamefont
  {Murata},\ and\ \citenamefont {Maeno}}]{Takatsu}%
  \BibitemOpen
  \bibfield  {author} {\bibinfo {author} {\bibfnamefont {H.}~\bibnamefont
  {Takatsu}}, \bibinfo {author} {\bibfnamefont {J.~J.}\ \bibnamefont
  {Ishikawa}}, \bibinfo {author} {\bibfnamefont {S.}~\bibnamefont {Yonezawa}},
  \bibinfo {author} {\bibfnamefont {H.}~\bibnamefont {Yoshino}}, \bibinfo
  {author} {\bibfnamefont {T.}~\bibnamefont {Shishidou}}, \bibinfo {author}
  {\bibfnamefont {T.}~\bibnamefont {Oguchi}}, \bibinfo {author} {\bibfnamefont
  {K.}~\bibnamefont {Murata}}, \ and\ \bibinfo {author} {\bibfnamefont
  {Y.}~\bibnamefont {Maeno}},\ }\bibfield  {title} {\enquote {\bibinfo {title}
  {Extremely large magnetoresistance in the nonmagnetic metal
  {${\mathrm{PdCoO}}_{2}$}},}\ }\href@noop {} {\bibfield  {journal} {\bibinfo
  {journal} {Phys. Rev. Lett.}\ }\textbf {\bibinfo {volume} {111}},\ \bibinfo
  {pages} {056601} (\bibinfo {year} {2013})}\BibitemShut {NoStop}%
\bibitem [{\citenamefont {Ong}\ \emph {et~al.}(2010)\citenamefont {Ong},
  \citenamefont {Singh},\ and\ \citenamefont {Wu}}]{Ong}%
  \BibitemOpen
  \bibfield  {author} {\bibinfo {author} {\bibfnamefont {K.~P.}\ \bibnamefont
  {Ong}}, \bibinfo {author} {\bibfnamefont {D.~J.}\ \bibnamefont {Singh}}, \
  and\ \bibinfo {author} {\bibfnamefont {P.}~\bibnamefont {Wu}},\ }\bibfield
  {title} {\enquote {\bibinfo {title} {Unusual transport and strongly
  anisotropic thermopower in {${\mathrm{PtCoO}}_{2}$} and
  {${\mathrm{PdCoO}}_{2}$}},}\ }\href@noop {} {\bibfield  {journal} {\bibinfo
  {journal} {Phys. Rev. Lett.}\ }\textbf {\bibinfo {volume} {104}},\ \bibinfo
  {pages} {176601} (\bibinfo {year} {2010})}\BibitemShut {NoStop}%
\bibitem [{\citenamefont {Eyert}\ \emph {et~al.}(2008)\citenamefont {Eyert},
  \citenamefont {Fr\'esard},\ and\ \citenamefont {Maignan}}]{Eyert}%
  \BibitemOpen
  \bibfield  {author} {\bibinfo {author} {\bibfnamefont {V.}~\bibnamefont
  {Eyert}}, \bibinfo {author} {\bibfnamefont {R.}~\bibnamefont {Fr\'esard}}, \
  and\ \bibinfo {author} {\bibfnamefont {A.}~\bibnamefont {Maignan}},\
  }\bibfield  {title} {\enquote {\bibinfo {title} {On the metallic conductivity
  of the delafossites {$\mathrm{PdCoO}_2$} and {$\mathrm{PtCoO}_2$}},}\
  }\href@noop {} {\bibfield  {journal} {\bibinfo  {journal} {Chem. Mater.}\
  }\textbf {\bibinfo {volume} {20}},\ \bibinfo {pages} {2370} (\bibinfo {year}
  {2008})}\BibitemShut {NoStop}%
\bibitem [{\citenamefont {Kikugawa}\ \emph {et~al.}(2015)\citenamefont
  {Kikugawa}, \citenamefont {Goswami}, \citenamefont {Kiswandhi}, \citenamefont
  {Choi}, \citenamefont {Graf}, \citenamefont {Baumbach}, \citenamefont
  {Brooks}, \citenamefont {Sugii}, \citenamefont {Iida}, \citenamefont
  {Nishio}, \citenamefont {Uji}, \citenamefont {Terashima}, \citenamefont
  {Rourke}, \citenamefont {Hussey}, \citenamefont {Takatsu}, \citenamefont
  {Yonezawa}, \citenamefont {Maeno},\ and\ \citenamefont {Balicas}}]{Kikugawa}%
  \BibitemOpen
  \bibfield  {author} {\bibinfo {author} {\bibfnamefont {N.}~\bibnamefont
  {Kikugawa}}, \bibinfo {author} {\bibfnamefont {P.}~\bibnamefont {Goswami}},
  \bibinfo {author} {\bibfnamefont {A.}~\bibnamefont {Kiswandhi}}, \bibinfo
  {author} {\bibfnamefont {E.~S.}\ \bibnamefont {Choi}}, \bibinfo {author}
  {\bibfnamefont {D.}~\bibnamefont {Graf}}, \bibinfo {author} {\bibfnamefont
  {R.~E.}\ \bibnamefont {Baumbach}}, \bibinfo {author} {\bibfnamefont {J.~S.}\
  \bibnamefont {Brooks}}, \bibinfo {author} {\bibfnamefont {K.}~\bibnamefont
  {Sugii}}, \bibinfo {author} {\bibfnamefont {Y.}~\bibnamefont {Iida}},
  \bibinfo {author} {\bibfnamefont {M.}~\bibnamefont {Nishio}}, \bibinfo
  {author} {\bibfnamefont {S.}~\bibnamefont {Uji}}, \bibinfo {author}
  {\bibfnamefont {T.}~\bibnamefont {Terashima}}, \bibinfo {author}
  {\bibfnamefont {P.~M.~C.}\ \bibnamefont {Rourke}}, \bibinfo {author}
  {\bibfnamefont {N.~E.}\ \bibnamefont {Hussey}}, \bibinfo {author}
  {\bibfnamefont {H.}~\bibnamefont {Takatsu}}, \bibinfo {author} {\bibfnamefont
  {S.}~\bibnamefont {Yonezawa}}, \bibinfo {author} {\bibfnamefont
  {Y.}~\bibnamefont {Maeno}}, \ and\ \bibinfo {author} {\bibfnamefont
  {L.}~\bibnamefont {Balicas}},\ }\bibfield  {title} {\enquote {\bibinfo
  {title} {Inter-planar coupling dependent magnetoresistivity in high purity
  layered metals},}\ }\href@noop {} {\bibfield  {journal} {\bibinfo  {journal}
  {arXiv:1412.5168v3}\ } (\bibinfo {year} {2015})}\BibitemShut {NoStop}%
\bibitem [{\citenamefont {Hicks}\ \emph
  {et~al.}(2012{\natexlab{a}})\citenamefont {Hicks}, \citenamefont {Gibbs},
  \citenamefont {Mackenzie}, \citenamefont {Takatsu}, \citenamefont {Maeno},\
  and\ \citenamefont {Yelland}}]{Hicks}%
  \BibitemOpen
  \bibfield  {author} {\bibinfo {author} {\bibfnamefont {C.~W.}\ \bibnamefont
  {Hicks}}, \bibinfo {author} {\bibfnamefont {A.~S.}\ \bibnamefont {Gibbs}},
  \bibinfo {author} {\bibfnamefont {A.~P.}\ \bibnamefont {Mackenzie}}, \bibinfo
  {author} {\bibfnamefont {H.}~\bibnamefont {Takatsu}}, \bibinfo {author}
  {\bibfnamefont {Y.}~\bibnamefont {Maeno}}, \ and\ \bibinfo {author}
  {\bibfnamefont {E.~A.}\ \bibnamefont {Yelland}},\ }\bibfield  {title}
  {\enquote {\bibinfo {title} {Quantum oscillations and high carrier mobility
  in the delafossite {${\mathrm{PdCoO}}_{2}$}},}\ }\href@noop {} {\bibfield
  {journal} {\bibinfo  {journal} {Phys. Rev. Lett.}\ }\textbf {\bibinfo
  {volume} {109}},\ \bibinfo {pages} {116401} (\bibinfo {year}
  {2012}{\natexlab{a}})}\BibitemShut {NoStop}%
\bibitem [{\citenamefont {Ziman}(1979)}]{Ziman}%
  \BibitemOpen
  \bibfield  {author} {\bibinfo {author} {\bibfnamefont {J.M.}\ \bibnamefont
  {Ziman}},\ }\href@noop {} {\emph {\bibinfo {title} {Principles of the
  {Theory} of {Solids}, 2nd ed.}}}\ (\bibinfo  {publisher} {Cambridge
  University Press},\ \bibinfo {address} {Cambridge},\ \bibinfo {year}
  {1979})\BibitemShut {NoStop}%
\bibitem [{\citenamefont {Grigoriev}(2010)}]{Grigoriev}%
  \BibitemOpen
  \bibfield  {author} {\bibinfo {author} {\bibfnamefont {P.~D.}\ \bibnamefont
  {Grigoriev}},\ }\bibfield  {title} {\enquote {\bibinfo {title} {Angular
  dependence of the {Fermi} surface cross-section area and magnetoresistance in
  quasi-two-dimensional metals},}\ }\href@noop {} {\bibfield  {journal}
  {\bibinfo  {journal} {Phys. Rev. B}\ }\textbf {\bibinfo {volume} {81}}
  (\bibinfo {year} {2010})}\BibitemShut {NoStop}%
\bibitem [{\citenamefont {MATLAB}(2014)}]{MATLAB}%
  \BibitemOpen
  \bibfield  {author} {\bibinfo {author} {\bibnamefont {MATLAB}},\ }\href@noop
  {} {\emph {\bibinfo {title} {version 8.4 (R2014b)}}}\ (\bibinfo  {publisher}
  {The MathWorks Inc.},\ \bibinfo {address} {Natick, Massachusetts},\ \bibinfo
  {year} {2014})\BibitemShut {NoStop}%
\bibitem [{\citenamefont {Kennett}\ and\ \citenamefont
  {McKenzie}(2007)}]{Kennett}%
  \BibitemOpen
  \bibfield  {author} {\bibinfo {author} {\bibfnamefont {M.~P.}\ \bibnamefont
  {Kennett}}\ and\ \bibinfo {author} {\bibfnamefont {R.~H.}\ \bibnamefont
  {McKenzie}},\ }\bibfield  {title} {\enquote {\bibinfo {title} {Sensitivity of
  the interlayer magnetoresistance of layered metals to intralayer
  anisotropies},}\ }\href@noop {} {\bibfield  {journal} {\bibinfo  {journal}
  {Phys. Rev. B}\ }\textbf {\bibinfo {volume} {76}},\ \bibinfo {pages} {054515}
  (\bibinfo {year} {2007})}\BibitemShut {NoStop}%
\bibitem [{\citenamefont {Yamaji}(1989)}]{Yamaji}%
  \BibitemOpen
  \bibfield  {author} {\bibinfo {author} {\bibfnamefont {K.}~\bibnamefont
  {Yamaji}},\ }\bibfield  {title} {\enquote {\bibinfo {title} {On the angle
  dependence of the magnetoresistance in quasi-two-dimensional organic
  superconductors},}\ }\href@noop {} {\bibfield  {journal} {\bibinfo  {journal}
  {J. Phys. Soc. Jpn.}\ }\textbf {\bibinfo {volume} {58}},\ \bibinfo {pages}
  {1520} (\bibinfo {year} {1989})}\BibitemShut {NoStop}%
\bibitem [{\citenamefont {Shockley}(1950)}]{Shockley}%
  \BibitemOpen
  \bibfield  {author} {\bibinfo {author} {\bibfnamefont {W.}~\bibnamefont
  {Shockley}},\ }\bibfield  {title} {\enquote {\bibinfo {title} {Effect of
  {Magnetic} {Fields} on {Conduction} ``{Tube} {Integrals}"},}\ }\href@noop {}
  {\bibfield  {journal} {\bibinfo  {journal} {Phys. Rev.}\ }\textbf {\bibinfo
  {volume} {79}},\ \bibinfo {pages} {191--192} (\bibinfo {year}
  {1950})}\BibitemShut {NoStop}%
\bibitem [{\citenamefont {Chambers}(1952)}]{Chambers}%
  \BibitemOpen
  \bibfield  {author} {\bibinfo {author} {\bibfnamefont {R.~G.}\ \bibnamefont
  {Chambers}},\ }\bibfield  {title} {\enquote {\bibinfo {title} {The {Kinetic}
  {Formulation} of {Conduction} {Problems}},}\ }\href@noop {} {\bibfield
  {journal} {\bibinfo  {journal} {Proc. Phys. Soc. A}\ }\textbf {\bibinfo
  {volume} {65}},\ \bibinfo {pages} {458} (\bibinfo {year} {1952})}\BibitemShut
  {NoStop}%
\bibitem [{\citenamefont {Blundell}\ \emph {et~al.}(1997)\citenamefont
  {Blundell}, \citenamefont {Ardavan},\ and\ \citenamefont
  {Singleton}}]{Blundell}%
  \BibitemOpen
  \bibfield  {author} {\bibinfo {author} {\bibfnamefont {S.~J.}\ \bibnamefont
  {Blundell}}, \bibinfo {author} {\bibfnamefont {A.}~\bibnamefont {Ardavan}}, \
  and\ \bibinfo {author} {\bibfnamefont {J.}~\bibnamefont {Singleton}},\
  }\bibfield  {title} {\enquote {\bibinfo {title} {Harmonics of the real-space
  velocity in cyclotron resonance experiments on organic metals},}\ }\href@noop
  {} {\bibfield  {journal} {\bibinfo  {journal} {Phys. Rev. B}\ }\textbf
  {\bibinfo {volume} {55}},\ \bibinfo {pages} {R6129--R6132} (\bibinfo {year}
  {1997})}\BibitemShut {NoStop}%
\bibitem [{\citenamefont {McKenzie}\ and\ \citenamefont
  {Moses}(1998)}]{McKenzie}%
  \BibitemOpen
  \bibfield  {author} {\bibinfo {author} {\bibfnamefont {R.~H.}\ \bibnamefont
  {McKenzie}}\ and\ \bibinfo {author} {\bibfnamefont {P.}~\bibnamefont
  {Moses}},\ }\bibfield  {title} {\enquote {\bibinfo {title} {Incoherent
  interlayer transport and angular-dependent magnetoresistance oscillations in
  layered metals},}\ }\href@noop {} {\bibfield  {journal} {\bibinfo  {journal}
  {Phys. Rev. Lett.}\ }\textbf {\bibinfo {volume} {81}},\ \bibinfo {pages}
  {4492} (\bibinfo {year} {1998})}\BibitemShut {NoStop}%
\bibitem [{\citenamefont {Yagi}\ \emph {et~al.}(1990)\citenamefont {Yagi},
  \citenamefont {Iye}, \citenamefont {Osada},\ and\ \citenamefont
  {Kagoshima}}]{Yagi}%
  \BibitemOpen
  \bibfield  {author} {\bibinfo {author} {\bibfnamefont {R.}~\bibnamefont
  {Yagi}}, \bibinfo {author} {\bibfnamefont {Y.}~\bibnamefont {Iye}}, \bibinfo
  {author} {\bibfnamefont {T.}~\bibnamefont {Osada}}, \ and\ \bibinfo {author}
  {\bibfnamefont {S.}~\bibnamefont {Kagoshima}},\ }\bibfield  {title} {\enquote
  {\bibinfo {title} {Semiclassical interpretation of the angular-dependent
  oscillatory magnetoresistance in quasi-two-dimensional systems},}\
  }\href@noop {} {\bibfield  {journal} {\bibinfo  {journal} {J. Phys. Soc.
  Jpn.}\ }\textbf {\bibinfo {volume} {59}},\ \bibinfo {pages} {3069--3072}
  (\bibinfo {year} {1990})}\BibitemShut {NoStop}%
\bibitem [{\citenamefont {Yakovenko}\ and\ \citenamefont
  {Cooper}(2006)}]{Yakovenko}%
  \BibitemOpen
  \bibfield  {author} {\bibinfo {author} {\bibfnamefont {V.~M.}\ \bibnamefont
  {Yakovenko}}\ and\ \bibinfo {author} {\bibfnamefont {B.~K.}\ \bibnamefont
  {Cooper}},\ }\bibfield  {title} {\enquote {\bibinfo {title} {Angular
  magnetoresistance oscillations in bilayers in tilted magnetic fields},}\
  }\href@noop {} {\bibfield  {journal} {\bibinfo  {journal} {Physica E}\
  }\textbf {\bibinfo {volume} {34}},\ \bibinfo {pages} {128--131} (\bibinfo
  {year} {2006})}\BibitemShut {NoStop}%
\bibitem [{\citenamefont {Kushwaha}\ \emph {et~al.}(2014)\citenamefont
  {Kushwaha}, \citenamefont {Moll}, \citenamefont {Nandi},\ and\ \citenamefont
  {Mackenzie}}]{Kushwaha}%
  \BibitemOpen
  \bibfield  {author} {\bibinfo {author} {\bibfnamefont {P.}~\bibnamefont
  {Kushwaha}}, \bibinfo {author} {\bibfnamefont {P.~J.~W.}\ \bibnamefont
  {Moll}}, \bibinfo {author} {\bibfnamefont {N.}~\bibnamefont {Nandi}}, \ and\
  \bibinfo {author} {\bibfnamefont {A.~P.}\ \bibnamefont {Mackenzie}},\
  }\bibfield  {title} {\enquote {\bibinfo {title} {Crystal growth, resistivity
  and {Hall} effect of the delafossite metal {PtCoO}$_2$},}\ }\href
  {http://arxiv.org/abs/1411.6162} {\bibfield  {journal} {\bibinfo  {journal}
  {arXiv:1411.6162}\ } (\bibinfo {year} {2014})}\BibitemShut {NoStop}%
\bibitem [{\citenamefont {Schofield}\ and\ \citenamefont
  {Cooper}(2000)}]{Schofield}%
  \BibitemOpen
  \bibfield  {author} {\bibinfo {author} {\bibfnamefont {A.~J.}\ \bibnamefont
  {Schofield}}\ and\ \bibinfo {author} {\bibfnamefont {J.~R.}\ \bibnamefont
  {Cooper}},\ }\bibfield  {title} {\enquote {\bibinfo {title} {Quasilinear
  magnetoresistance in an almost two-dimensional band structure},}\ }\href@noop
  {} {\bibfield  {journal} {\bibinfo  {journal} {Phys. Rev. B}\ }\textbf
  {\bibinfo {volume} {62}},\ \bibinfo {pages} {10779} (\bibinfo {year}
  {2000})}\BibitemShut {NoStop}%
\bibitem [{\citenamefont {Abramowitz}\ and\ \citenamefont
  {Stegun}(1964)}]{Abramowitz}%
  \BibitemOpen
  \bibfield  {author} {\bibinfo {author} {\bibfnamefont {M.}~\bibnamefont
  {Abramowitz}}\ and\ \bibinfo {author} {\bibfnamefont {I.~A.}\ \bibnamefont
  {Stegun}},\ }\href@noop {} {\emph {\bibinfo {title} {Handbook of Mathematical
  Functions with Formulas, Graphs, and Mathematical Tables}}}\ (\bibinfo
  {publisher} {Dover},\ \bibinfo {address} {New York},\ \bibinfo {year}
  {1964})\BibitemShut {NoStop}%
\bibitem [{\citenamefont {Hicks}\ \emph
  {et~al.}(2012{\natexlab{b}})\citenamefont {Hicks}, \citenamefont {Gibbs},
  \citenamefont {Mackenzie}, \citenamefont {Takatsu}, \citenamefont {Maeno},\
  and\ \citenamefont {Yelland}}]{HicksSuppl}%
  \BibitemOpen
  \bibfield  {author} {\bibinfo {author} {\bibfnamefont {C.~W.}\ \bibnamefont
  {Hicks}}, \bibinfo {author} {\bibfnamefont {A.~S.}\ \bibnamefont {Gibbs}},
  \bibinfo {author} {\bibfnamefont {A.~P.}\ \bibnamefont {Mackenzie}}, \bibinfo
  {author} {\bibfnamefont {H.}~\bibnamefont {Takatsu}}, \bibinfo {author}
  {\bibfnamefont {Y.}~\bibnamefont {Maeno}}, \ and\ \bibinfo {author}
  {\bibfnamefont {E.~A.}\ \bibnamefont {Yelland}},\ }\bibfield  {title}
  {\enquote {\bibinfo {title} {Supplementary {Material} of `{Quantum}
  oscillations and high carrier mobility in the delafossite
  {${\mathrm{PdCoO}}_{2}$}'},}\ }\href@noop {} {\bibfield  {journal} {\bibinfo
  {journal} {Phys. Rev. Lett.}\ }\textbf {\bibinfo {volume} {109}},\ \bibinfo
  {pages} {116401} (\bibinfo {year} {2012}{\natexlab{b}})}\BibitemShut
  {NoStop}%
\end{thebibliography}%


\newpage

\section{APPENDICES}
\subsection{Linearised Boltzmann transport equation}  \label{sec:BoltzmannAppendix}
The oscillations in AMRO are semi-classical in nature, and are due to the formation of cyclotron orbits on the FS and the changing area enclosed by them, as shown by Yamaji \cite{Yamaji}. To be able to find the conductivity and thus the resistivity, we need an equation that describes the evolution of the quasiparticle distribution in our material semi-classically - the linearised Boltzmann transport equation\cite{Ziman}. To arrive at this, firstly we define $f_\mathbf{k}(\mathbf{r})$ as the steady state distribution of charged fermions under the influence of scattering, diffusion and electromagnetic forces. We can then assert that
\begin{equation} \label{eq:DistributionEqn} \frac{\partial f_\mathbf{k}}{\partial t}_{\text{total}} = \frac{\partial f_\mathbf{k}}{\partial t}_{\text{scattering}} + \frac{\partial f_\mathbf{k}}{\partial t}_{\text{diffusion}} + \frac{\partial f_\mathbf{k}}{\partial t}_{\text{EM forces}} = 0 . \end{equation}
At constant temperature, we can neglect the diffusion term. The scattering term is approximated using the relaxation time approximation - we define $g_\mathbf{k} = f_\mathbf{k} - f_\mathbf{k}^0$, where $f_\mathbf{k}^0$ is the equilibrium distribution, and a scattering time $\tau$, and write the scattering term as
\begin{equation} \frac{\partial f_\mathbf{k}}{\partial t}_{\text{scattering}} = -\frac{g_\mathbf{k}}{\tau} . \end{equation}
The electromagnetic term is written as
\begin{equation} \frac{\partial f_\mathbf{k}}{\partial t}_{\text{EM fields}} = \frac{\partial f_\mathbf{k}}{\partial \mathbf{k}} \frac{\partial \mathbf{k}}{\partial t} = - \frac{e}{\hbar}(\mathbf{E} + \mathbf{v}\times\mathbf{B}) \cdot \left( \frac{\partial g_\mathbf{k}}{\partial \mathbf{k}} + \frac{\partial f_\mathbf{k}^0}{\partial \mathbf{k}} \right) . \end{equation}
Substituting these expressions into Eq. \eqref{eq:DistributionEqn} and using $\frac{\partial f_\mathbf{k}^0}{\partial \mathbf{k}} = \frac{\partial f_\mathbf{k}^0}{\partial \varepsilon} \frac{\partial \varepsilon}{\partial \mathbf{k}} = \frac{\partial f_\mathbf{k}^0}{\partial \varepsilon} \hbar \mathbf{v}$, we get
\begin{equation} - \frac{e}{\hbar} (\mathbf{E} + \mathbf{v}\times \mathbf{B}) \cdot \frac{\partial g_\mathbf{k}}{\partial \mathbf{k}} - e (\mathbf{E} + \mathbf{v}\times \mathbf{B})\cdot \mathbf{v} \frac{\partial f_\mathbf{k}^0}{\partial \varepsilon} = \frac{g_\mathbf{k}}{\tau} . \end{equation}
Applying $\mathbf{v}\cdot(\mathbf{v}\times\mathbf{B})=0$ and neglecting the $\mathbf{E}\cdot \frac{\partial g_\mathbf{k}}{\partial \mathbf{k}}$, as it is an quadratic deviation from Ohm's Law and we wish to linearise the equation, we get
\begin{equation} e \mathbf{E}\cdot \mathbf{v} \left( - \frac{\partial f_\mathbf{k}^0}{\partial \varepsilon} \right) = \frac{g_\mathbf{k}}{\tau} + \frac{e}{\hbar}(\mathbf{v}\times\mathbf{B}) \cdot \frac{\partial g_\mathbf{k}}{\partial \mathbf{k}} . \end{equation}

\subsection{Shockley-Chambers tube integral}  \label{sec:SCTIAppendix}
We now need to solve the linearised Boltzmann transport equation to obtain the conductivity as an integral equation. If we only have a magnetic field acting on the system, the Lorentz force is $\mathbf{F} = \frac{\partial \mathbf{k}}{\partial t} = - \frac{e}{\hbar} \mathbf{v}\times\mathbf{B}$. Substituting this into Eq. \eqref{eq:Boltzmannequation}, and defining the phase angle round the orbit $\alpha$ as $\partial \alpha = \omega_c \partial t$, where $\omega_c = \frac{eB}{m^*}$, we obtain
\begin{equation} e\mathbf{E}\cdot\mathbf{v}\left( - \frac{\partial f_\mathbf{k}^0}{\partial \varepsilon} \right) = \frac{g_\mathbf{k}}{\tau} + \omega_c \frac{\partial g_\mathbf{k}}{\partial \alpha} . \end{equation}
The right hand side of this equation can be written as $ \omega_c e^{-\frac{\alpha}{\omega_c \tau}} \frac{\partial}{\partial \alpha} \left( e^{\frac{\alpha}{\omega_c \tau}} g_\mathbf{k} \right) $, allowing us to obtain (using Einstein summation notation):
\begin{equation} \begin{aligned} \label{eq:gkequation} g_\mathbf{k} &= \left( - \frac{\partial f_\mathbf{k}^0}{\partial \varepsilon} \right) e^{-\frac{\alpha}{\omega_c \tau}} \int_{-\infty}^{\alpha} \frac{e}{\omega_c} e^{\frac{\alpha'}{\omega_c \tau}} E_j v_j (\alpha') d\alpha' \\
& = \left( - \frac{\partial f_\mathbf{k}^0}{\partial \varepsilon} \right) \int_0^{\infty} \frac{e}{\omega_c} e^{\frac{\alpha''}{\omega_c \tau}} E_j v_j (\alpha - \alpha'') d\alpha'' . \end{aligned} \end{equation}

Now we need an expression for $d^3 \mathbf{k}$ to complete the integration. If we define $k_B$ as parallel to $\mathbf{B}$, $k_\perp$ in the direction of increasing $\alpha$ and $k_\parallel$ along the radius of the orbit, we can write $d^3 \mathbf{k} = dk_B dk_\perp dk_\parallel$. We can express $dk_\parallel$ in terms of the tangential velocity $v_\perp$ using $\frac{\partial \mathbf{k}}{\partial t} = \frac{e}{\hbar} \mathbf{v} \times \mathbf{B}$ and $d\alpha = \omega_c dt$:
\begin{equation} dk_\parallel = \frac{ev_\perp B}{\hbar} dt = \frac{v_\perp m^*}{\hbar} d \alpha .\end{equation}
We can also write the differential of the energy as
\begin{equation} d \varepsilon = \hbar \frac{1}{\hbar} \frac{\partial \varepsilon}{\partial k_\perp} dk_\perp = \hbar v_\perp dk_\perp . \end{equation}
Putting all of this together, we obtain the current density $ J_i = \frac{1}{4\pi^3} \int e v_i g_\mathbf{k} d^3 \mathbf{k} = \sigma_{ij} E_j $ as
\begin{equation} J_i = \frac{e}{4\pi^3} \int d\varepsilon \int dk_B \int d\alpha \frac{m^*}{\hbar^2} v_i (\alpha) g_\mathbf{k} . \end{equation}
Substituting in the expression for $g_\mathbf{k}$, and using the fact that, as $f_\mathbf{k}^0$ is the Fermi-Dirac distribution and $T\ll T_F\simeq10^5$~K, $(- \frac{\partial f_\mathbf{k}^0}{\partial \varepsilon}) = \delta(\varepsilon - \varepsilon_F)$, we obtain
\begin{equation} \begin{aligned} \sigma_{ij} = \frac{e^2}{4\pi^3\hbar^2} &\int dk_B \int_0^{2\pi} d\alpha \int_0^{\infty} d\alpha'' \\ &v_i (\alpha) v_j (\alpha - \alpha'') \frac{m^*}{\omega_c} e^{-\frac{\alpha''}{\omega_c \tau}} . \end{aligned} \end{equation}

This expression can be generalised to include open orbits, as shown in Eq. \eqref{eq:GeneralisedSCTI} in the main text, to give:
\begin{equation} \sigma_{ij} = \frac{e^2}{4\pi^3} \int d^3 \mathbf{k} \left(-\frac{\partial f_\mathbf{k}^0}{\partial \varepsilon}\right) v_i (\mathbf{k},0) \int_{-\infty}^0 v_j (\mathbf{k},t) e^{\frac{t}{\tau}} dt . \end{equation}

\subsection{AMRO in-plane integral}  \label{sec:InPlaneAppendix}
From Eq. \eqref{eq:GeneralisedSCTI}, one can obtain the equation (as in Ref. \onlinecite{Schofield})
\begin{equation} \sigma_{ij} = \frac{e^2}{4\pi^3} \oint \frac{dS}{\hbar |\mathbf{v}|} \int_0^\infty v_i (0) v_j (t) e^{-\frac{t}{\tau}} dt \end{equation}
where $dS$ is an area element of the FS. We can neglect the small closed orbits that will form on the sides of the FS, as they will give a very small contribution to the conductivity. The velocities depend on time through $k_z$, which, if we take the $z$-component of the Lorentz force, is given by
\begin{equation} \hbar \frac{dk_z}{dt} = -ev_F B \sin \theta \end{equation}
where $\theta = \mathbf{v}\wedge\mathbf{B} = \psi - \phi$. $\psi$ is the azimuthal angle that will be integrated over, and $\phi$ is the azimuthal angle of the magnetic field. Making the approximation that $k_F \simeq k_{00}$, this equation is solved by $k_z(t) = - \frac{ek_{00}B}{m^*} t \sin \theta  +k_z (0)$. $k_z(0)$ will be one of the variables of integration. $|\mathbf{v}| = v_F = \frac{\hbar}{m^*} k_F$, so putting all of this together, we obtain
\begin{equation} \sigma_{ij} = \frac{e^2}{4\pi^3} \frac{m^*}{\hbar^2} \int_{-\frac{3\pi}{c}}^{\frac{3\pi}{c}} dk_z \int_0^{2\pi} d\psi \int_0^{\infty} dt \frac{v_i (0) v_j (t) e^{-\frac{t}{\tau}}}{k_F (0)} . \end{equation}

In order to compute this using MATLAB\cite{MATLAB}, the variable change $x = k_z (t)$ was made, giving the integral
\begin{equation} \begin{aligned} \sigma_{ij} = \frac{e^2}{4\pi^3} \frac{m^*}{\hbar^2} &\int_{-\frac{3\pi}{c}}^{\frac{3\pi}{c}} dk_z \int_0^{2\pi} d\psi \int_{-\infty}^{k_z} dx \frac{v_i (k_z,\psi) v_j (x,\psi)}{\omega k_F (k_z,\psi)} \\ &e^{-\frac{k_z}{\omega \tau \sin \theta}} e^{\frac{x}{\omega \tau \sin \theta}} \end{aligned} \end{equation}
where $\omega = \frac{ek_{00} B}{m^*}$.

\subsection{AMRO out-of-plane integral}  \label{sec:OutOfPlaneAppendix}
Starting from Eq. \eqref{eq:SCTI}, we can project $dk_B$ onto the c-axis, giving $dk_B = dk_0 \cos \theta$. In addition to this, we can imagine rotating the whole co-ordinate system by $\theta$, and consider the orbit due to a field of $B$ at angle $\theta$ to be due to a field of $B \cos \theta$ along the $c$-axis, as long as we still enter the correct value of $k_z$ into $k_F$. This has the effect of taking $\omega_c \to \omega_c \cos \theta$, and $k_z= k_0 - k_F (k_0, \alpha) \cos \alpha \tan \theta$. Putting all of this together, we obtain the equation
\begin{equation} \begin{aligned} \sigma_{ij} &= \frac{e^2 m^*}{4\pi^3 \hbar^2 \omega_c} \int_{-\frac{3\pi}{c}}^{\frac{3\pi}{c}} dk_0 \int_0^{2\pi} d\alpha \int_0^{\infty} d\alpha'' \\ &e^{-\frac{\alpha''}{\omega_c \tau \cos \theta}} v_i (k_0 - k_F(k_0 , \alpha + \phi) \cos \alpha \tan \theta , \alpha) \\ &v_j (k_0 - k_F(k_0 , \alpha - \alpha'' + \phi) \cos (\alpha - \alpha'') \tan \theta , \alpha - \alpha'') . \end{aligned} \end{equation}
Now, we write the velocities as Fourier series \cite{Analytis,Blundell}, defined as
\begin{equation} \begin{aligned} &v_i ( k_0 - k_F ( k_0 , \alpha) \cos \alpha \tan \theta , \alpha ) = \\ &\sum_{n=0}^{\infty} a_n \cos n\alpha + b_n \sin n\alpha , \end{aligned} \end{equation}
\begin{equation} \begin{aligned} &v_j ( k_0 - k_F ( k_0 , \alpha - \alpha'') \cos (\alpha - \alpha'') \tan \theta , \alpha - \alpha'') = \\ &\sum_{n=0}^{\infty} c_n \cos n(\alpha - \alpha'') + d_n \sin n(\alpha - \alpha''). \end{aligned}\end{equation}
We then expand out $\cos n(\alpha - \alpha'') = \cos n \alpha \cos n \alpha'' + \sin n \alpha \sin n \alpha''$ and $\sin n(\alpha - \alpha'') = \sin n \alpha \cos n \alpha'' - \cos n \alpha \sin n \alpha''$. The $\alpha''$ integral can now be completed, using
\begin{equation} \int_0^\infty \cos n \alpha'' e^{-\frac{\alpha''}{\omega_c \tau \cos \theta}} d \alpha'' = \frac{\omega_c \tau \cos \theta}{1+(\omega_c \tau \cos \theta)^2 n^2} , \end{equation}
\begin{equation} \int_0^\infty \sin n \alpha'' e^{-\frac{\alpha''}{\omega_c \tau \cos \theta}} d \alpha'' = \frac{n (\omega_c \tau \cos \theta)^2}{1+(\omega_c \tau \cos \theta)^2 n^2} . \end{equation}
This leaves us with the integral
\begin{equation} \begin{aligned} \sigma_{ij} &= \frac{e^2 m^*}{4\pi^3 \hbar^2 \omega_c} \int_{-\frac{3\pi}{c}}^{\frac{3\pi}{c}} dk_0 \sum_{n,m=0}^\infty \int_0^{2\pi} d\alpha \\ &\left(a_m \cos m\alpha + b_m \sin m\alpha \right) \left(c_n \left[ \frac{\omega_c \tau \cos \theta \cos n \alpha}{1+(\omega_c \tau \cos \theta)^2 n^2} + \right. \right. \\ & \left. \frac{n (\omega_c \tau \cos \theta)^2 \sin n \alpha}{1+(\omega_c \tau \cos \theta)^2 n^2} \right] + d_n \left[ \frac{\omega_c \tau \cos \theta \sin n \alpha}{1+(\omega_c \tau \cos \theta)^2 n^2} - \right. \\ & \left. \left. \frac{n (\omega_c \tau \cos \theta)^2 \cos n \alpha}{1+(\omega_c \tau \cos \theta)^2 n^2} \right] \right) . \end{aligned} \end{equation}
Finally, we can complete the integrals over $\alpha$ by using (for $m$ or $n \neq 0$):
\begin{equation} \int_0^{2\pi} \cos m \alpha \cos n \alpha = \pi \delta_{mn} ,\end{equation}
\begin{equation} \int_0^{2\pi} \sin m \alpha \sin n \alpha = \pi \delta_{mn} , \end{equation}
\begin{equation} \int_0^{2\pi} \sin m \alpha \cos n \alpha = 0 ,\end{equation}
and for $m,n=0$
\begin{equation} \int_0^{2\pi} \cos m \alpha \cos n \alpha = 2\pi , \end{equation}
\begin{equation} \int_0^{2\pi} \sin m \alpha \sin n \alpha = 0 . \end{equation}
This gives, finally, the equation
\begin{equation} \begin{aligned} &\sigma_{ij} = \frac{e^2}{4\pi^3\hbar^2} \frac{m^*}{\omega_c} \int_{-\frac{3\pi}{c}}^{\frac{3\pi}{c}} dk_0 \, a_0 c_0 + \\ &\frac{1}{2} \sum_{n=1}^\infty \left[ \frac{a_n c_n + b_n d_n}{1+(\omega_c \tau_0 \cos \theta)^2 n^2} - \frac{ (a_n d_n - b_n c_n) n\omega_c \tau_0 \cos \theta}{1 + (\omega_c \tau_0 \cos \theta)^2 n^2} \right] . \end{aligned} \end{equation}

\subsection{Velocities} \label{sec:VelocityAppendix}
The FS is effectively defined by an equation, in cylindrical co-ordinates, of the form $|\mathbf{k}|=f(\psi,k_z)$. The Fermi velocity is defined by $v_F = \frac{1}{\hbar} \frac{\partial \varepsilon_F}{\partial \mathbf{k}} = \frac{\hbar k_F}{m^*} \frac{\partial k_F}{\partial \mathbf{k}}$. This means that the velocity is perpendicular to the FS at all times. Calculating the derivative, we find that the velocity is in the direction $\hat{\mathbf{k}} - \frac{1}{k_F} \frac{\partial k_F}{\partial \psi} \hat{\boldsymbol{\psi}} -\frac{\partial k_F}{\partial k_z} \hat{\mathbf{k}_z}$. This tells us that $v_z = \frac{\hbar k_F}{m^*}\frac{\partial k_F}{\partial k_z}$, so the remaining velocity is given by $\sqrt{\left( \frac{\hbar k_F}{m^*} \right)^2 - v_z^2}$. If we then define the angle between the in-plane velocity and the x-axis as $\gamma$ (the angle between the radial direction and the x-axis is $\psi$), we can easily see that $\tan(\psi-\gamma) = - \frac{1}{k_F} \frac{\partial k_F}{\partial \psi}$. Rearranging this, we obtain $\gamma = \psi + \arctan \left( \frac{1}{k_F} \frac{\partial k_F}{\partial \psi} \right)$. It is then simple to see that
\begin{equation} v_x = \cos \gamma \sqrt{ \left( \frac{\hbar k_F}{m^*} \right)^2 - v_z^2} , \end{equation}
\begin{equation} v_y = \sin \gamma  \sqrt{ \left( \frac{\hbar k_F}{m^*} \right)^2 - v_z^2} . \end{equation}
We can now take advantage of the fact that $\frac{1}{k_F} \frac{\partial k_F}{\partial \psi}$ is small  (at maximum $\simeq0.48$ for $k_{1,3}=0.1$~$\text{\r{A}}^{-1}$), allowing us to neglect it, making $\gamma \simeq \psi$.  As $\frac{\partial k_F}{\partial k_z}\simeq k_{1,3} \ll 1$ at maximum for large $k_{1,3}$, $v_z^2$ is much less than $\left( \frac{\hbar k_F}{m^*} \right)^2$, allowing us to neglect it too, leading to the simple expressions
\begin{equation} v_x = \cos \psi \frac{\hbar k_F}{m^*} , \end{equation}
\begin{equation} v_y = \sin \psi \frac{\hbar k_F}{m^*} . \end{equation}

\subsection{Full matrix and simplified methods of calculating AMRO}

\begin{figure}[htbp]
\centering
\includegraphics[width=0.5\textwidth]{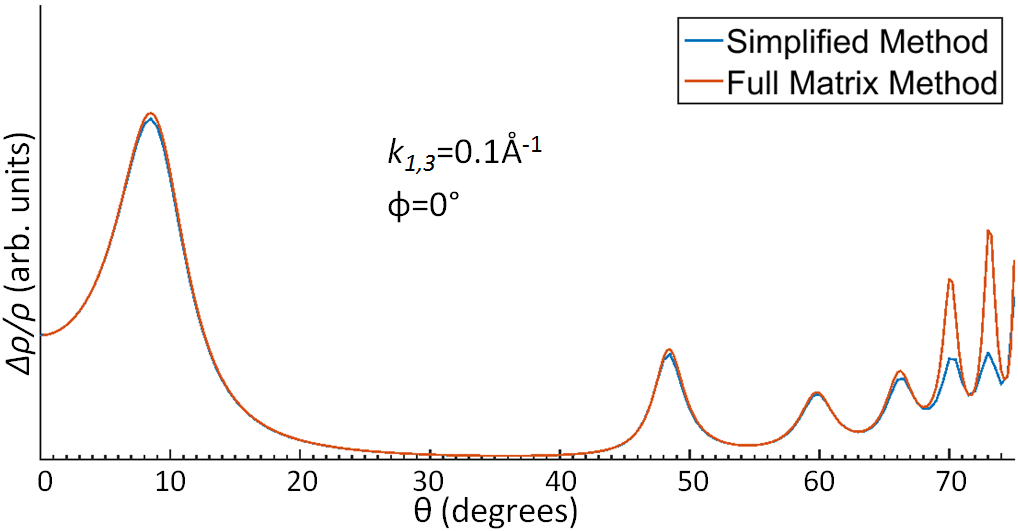}
\caption{(colour online) Simulated AMRO calculated using the full matrix method, where the entire conductivity matrix is calculated and then inverted to obtain $\rho_{zz}$, and the simplified method, where $\rho_{zz}=\frac{1}{\sigma_{zz}}$ is used. The positions of the peaks and form of the AMRO produced by the two methods agree very well, suggesting the less computationally expensive and error-prone simplified method can be used instead of the full matrix method. $k_{1,3}=0.1$~$\text{\r{A}}^{-1}$, $\phi=0^\circ$, $\tau=10$~ps and $B=25$~T in both simulations.} \label{fig:ComparingSimplifiedCalcs}
\end{figure}

\subsection{Yamaji angles and mapping the FS}

\begin{figure}[t]
\centering
\includegraphics[width=0.5\textwidth]{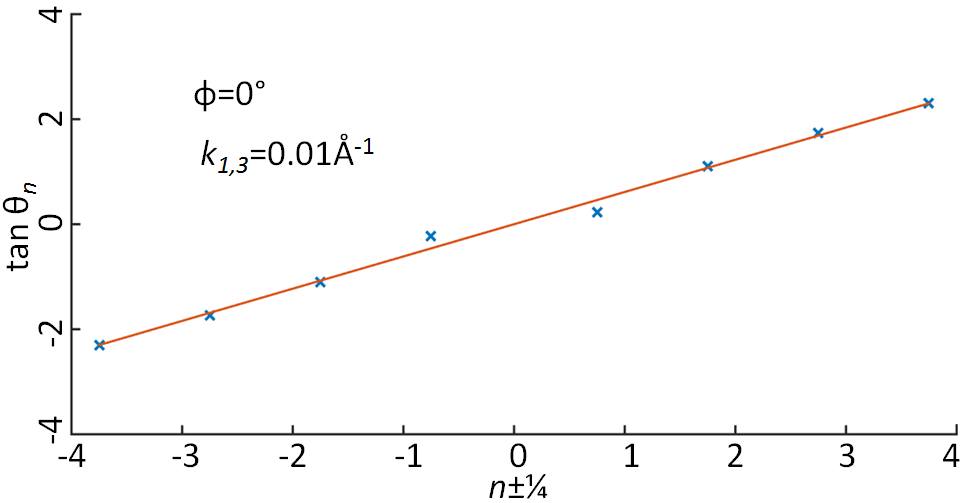}
\caption{(colour online) Fitting of AMRO peaks to the Yamaji angle formula, as shown in Eq. \eqref{eq:kparallelequation}, for $k_{1,3}=0.01$~$\text{\r{A}}^{-1}$ and $\phi=0^\circ$. The gradient of the line gives $k_\parallel=0.851\pm0.006$~$\text{\r{A}}^{-1}$. The central peaks (those with the smallest value of $|n\pm\frac{1}{4}|$) can clearly be seen to be slightly off the straight line defined by the other peaks, as would be expected due to the breakdown of the approximation that $\mu\gg1$. The fit shown does not use the central peaks - if these peaks are included, the error in the gradient increases by an order of magnitude, but the line is indistinguishable from the fit shown.} \label{fig:KParFitting}
\end{figure}

In order to map the FS using AMRO, we need a way of extracting $k_{\parallel}$ from the data for different values of the azimuthal angle, $\phi$. In this work the chosen parameters were $B=25\text{T}, \tau = 10\text{ps}, m^* = 1.5m_e$ and thus $\omega_c \tau$ is large enough to neglect terms other than $J_0 (\mu)$ in the Bessel function sum. This implies we will get minima in $\sigma_{zz}$, and thus maxima in $\rho_{zz}$, at the zeros of $J_0 (\mu)$. For $|\mu| \gg 1$, we can use the large argument expansion of the zeroth order Bessel function \cite{Abramowitz}:
\begin{equation} J_0 (\mu) \simeq \sqrt{\frac{2}{\pi \mu}} \left\{
\begin{array}{lr}
\cos \left( \mu - \frac{\pi}{4} \right) & \mu>0\\
\cos \left( \mu +\frac{\pi}{4} \right) & \mu<0
\end{array} \right. . \end{equation}
This has zeros at $\mu = n \pi + \frac{\pi}{2} \pm \frac{\pi}{4}$, where $+$ is for $\mu<0$ and $-$ is for $\mu>0$. We can then index each of the peaks in the AMRO found with an integer $n$, so the $\theta$ value of the peak is given by $\theta_n$. $n<0$ if $\theta_n<0$, and vice versa, and $|n|$ grows as $|\theta_n|$ grows. If we index the peaks in this way, the results above tell us that we should be able to plot $\tan \theta_n$ against $n \pm \frac{1}{4}$ and fit it to the equation
\begin{equation} \label{eq:kparallelequation} \frac{ck_\parallel}{3} \tan \theta_n = \pi (n \pm \frac{1}{4}) + C(\phi)\end{equation}
where the $ \pm $ signs are taken as described above, and $C(\phi)$ is a constant dependent on $\phi$. This provides us with a way of extracting $k_\parallel$ and mapping the FS, as required. By considering the orbits of electrons in magnetic field, we can explain the origin of these peaks. At certain values of $\theta$, all orbits have the same area,  resulting in the z-component of the velocity averaging to zero and therefore a peak in magnetoresistance. Yamaji showed that these special values of $\theta$ precisely correspond to the peaks already mentioned \cite{Yamaji}, and are thus often called Yamaji angles\cite{Lewin}.

\subsection{Simulation of AMROs for a tetragonal Fermi surface}

\begin{figure}[t]
\centering
\includegraphics[width=0.5\textwidth]{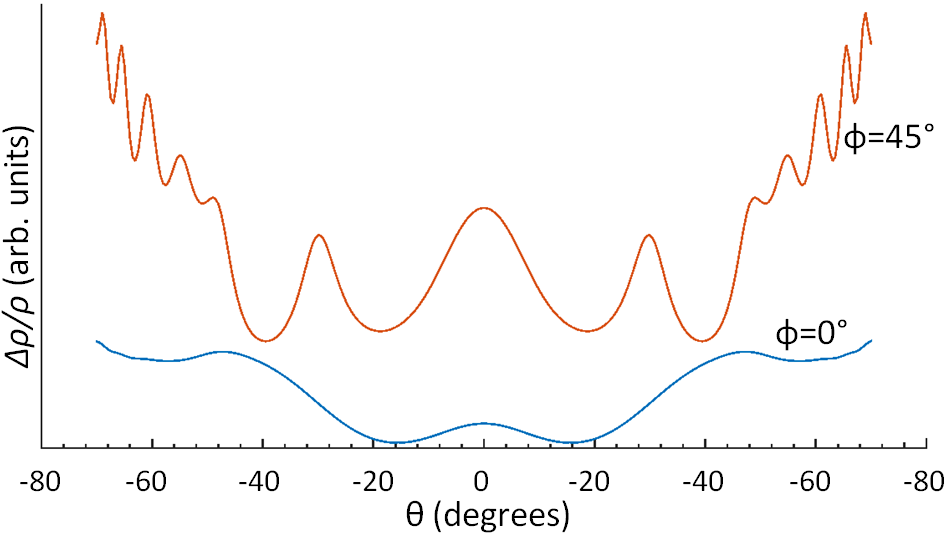}
\caption{(colour online) Calculated AMROs of a layered tetragonal material, with a FS similar to that of Tl$_2$Ba$_2$CuO$_{6+\delta}$: $k_F=k_{0,0}+k_{0,4}\cos 4\psi + k_{1,2}\cos \kappa \sin 2\psi + k_{1,6} \cos \kappa \sin 6\psi + k_{1,10} \cos \kappa \sin 10\psi$. Here $k_{0,0}=0.729$, $k_{0,4}=-0.0219$, $k_{1,2}=0.0031$, $k_{1,6}=0.00217$ and $k_{1,10}=-0.00093$~$\text{\r{A}}^{-1}$, with $B=25$~T, $\phi=0^\circ$ and $45^\circ$, $\tau=0.5$~ps and $m^*=4.1m_e$, similar to previous work\cite{Analytis}.} \label{fig:TetragonalAMRO}
\end{figure}

To test the robustness of our approach we have also
performed calculations for a layered tetragonal system, Tl$_2$Ba$_2$CuO$_{6+\delta}$,
 shown in Fig. \ref{fig:TetragonalAMRO}, which has strong AMRO features at low angles and agree qualitatively with available experiments \cite{Analytis}. This strengthens the generality of our method, its approximations and its applicability, which can be extended to other quasi-2D materials of various lattice symmetries.

\subsection{Tight-binding model}

 The terms of the Fermi surface expansion used in this work, given by Eq. \eqref{eq:kfexpansion}, can be linked to a tight-binding expansion involving both inter- and intra-plane nearest neighbour and next-nearest neighbour hopping terms \cite{Takatsu,HicksSuppl}. From a given tight-binding model and Fermi energy, a Fermi surface can be calculated that corresponds to that given by Eq. \eqref{eq:kfexpansion}. Intra-plane next-nearest neighbour hopping as well as inter-plane hopping affect the value of $k_{1,3}$ in the expansion, whilst the values of the in-plane next-nearest neighbour transfer integral and the Fermi energy affect the overall hexagonal shape. In previous tight-binding model calculations, the transfer integrals $t_{nn},t_{nnn},t_{z},t_{zz}$ were given values of approximately $1.0,-0.23,0.042,0.011$~eV, representing intra-plane nearest neighbour, in-plane next-nearest neighbour, inter-plane nearest neighbour and inter-plane next-nearest neighbour coupling respectively. The Fermi energy $E_F$ was chosen as $0.22$~eV \cite{HicksSuppl}. One possible tight-binding expression for the energy is of the form \cite{Takatsu}
\begin{equation} \begin{aligned} \varepsilon(\mathbf{k}) &= -2t_{nn}\{ \cos(\mathbf{k}\cdot\mathbf{a}) + \cos(\mathbf{k}\cdot\mathbf{b}) + \cos(-\mathbf{k}\cdot(\mathbf{a}+\mathbf{b}))\} \\& - \frac{2t_{zz}}{3} \cos(\mathbf{k}\cdot\mathbf{c}) - \sqrt{3}t_{nnn} \{\cos^2 (\mathbf{k}\cdot\mathbf{a}) + \cos^2 (\mathbf{k}\cdot\mathbf{b}) \\&+ \cos^2 (-\mathbf{k}\cdot(\mathbf{a}+\mathbf{b}))\} . \end{aligned} \end{equation}
Although using a tight-binding model gives access to more information than simply the geometry of the Fermi surface, we choose to work directly from the Fermi surface expansion because it gives more direct insight into what is actually observed experimentally - the shape of the Fermi surface itself \cite{Bergemann}. 

\subsection{Symmetry operations}

The symmetry operations used to construct the Fermi surface expansion in Section II can be described more intuitively in Cartesian co-ordinates as follows:
\begin{enumerate}
\item A screw symmetry, composed of a rotation about the origin by $60^\circ$ and a translation in the $z$-direction by half a reciprocal lattice vector.
\item A rotation symmetry about the origin by $120^\circ$.
\item Inversion through the origin.
\item Reflection in the plane $x=0$.
\item Reflection in the plane $y=\sqrt{3}x$, i.e. a line at a $60^\circ$ angle to the $x$-axis.
\end{enumerate}
The screw symmetry is responsible for the lack of a $k_{0,1}$ term, present in Ref. \onlinecite{Hicks}, in the Fermi surface expansion used in this work, which is given by Eq. \eqref{eq:kfexpansion}. 

\end{document}